\documentclass[traditabstract]{aa} 
\usepackage{graphicx}
\usepackage{txfonts}
\usepackage[T1]{fontenc}
\usepackage{subfigure}
\usepackage{subfig}
\usepackage{amssymb}
\usepackage{epstopdf}
\usepackage{natbib}
\bibpunct{(}{)}{;}{a}{}{,} 
\usepackage{url}

\begin{document}

\def\reac#1#2{\begin{equation}\label{r:r#1}\ce{#2}\end{equation}}
\def\reacnn#1{\[\ce{#1}\]}
\def\refreac#1{(\ref{r:r#1})}
\def\abS{\ensuremath{[\ce{S}]_{\rm tot}}}
\def\ab#1{\ensuremath{[\ce{#1}]}}
\def\rop{\ensuremath{\rm o/p}}
\def\tkin{\ensuremath{T_{\rm K}}}
\def\tspin{\ensuremath{T_{\rm spin}}}

\titlerunning {Ortho-to-para ratio of NH$_2$}
\title{Ortho-to-para ratio of NH$_2$}
\subtitle{\emph{Herschel}\thanks{\emph{Herschel} is an ESA space observatory with science instruments provided by European-led  Principal Investigator consortia and with important participation from NASA.}-HIFI observations of ortho- and
para-NH$_2$ rotational transitions towards  W31C, W49N, W51, and G34.3+0.1}
\authorrunning{C.M.~Persson et~al.}   
  \author{C.M.~Persson     
          \inst{1},
          A.O.H.~Olofsson\inst{1}, 
          R.~Le~Gal\inst{2},
          E.S.~Wirstr\"om\inst{1},
      G.~E.~Hassel\inst{3},  
                            E.~Herbst\inst{2},
            M.~Olberg\inst{1},
          A.~Faure\inst{4}, 
        P.~Hily-Blant\inst{4}, 
         J.H.~Black\inst{1},
         M.~Gerin\inst{5},
            D.~Lis\inst{6}, 
F.~Wyrowski\inst{7}   
          }   
   \offprints{carina.persson@chalmers.se}
   \institute{Chalmers University of Technology, Department of Earth and Space Sciences, Onsala Space Observatory,  SE-439 92 Onsala, Sweden.
    \email{\url{carina.persson@chalmers.se}} 
    \and Department of Chemistry, University of Virginia, McCormick Road, Charlottesville, VA 22904, USA 
    \and Department of Physics \& Astronomy, Siena College, Loudonville, NY  12211,   USA  
    \and Université Grenoble Alpes and CNRS, IPAG, F-38000 Grenoble, France 
\and LERMA, Observatoire de Paris, PSL Research University, CNRS, Sorbonne Universités, UPMC Univ. Paris 06, École normale supérieure, F-75005, Paris, France 
\and LERMA, Observatoire de Paris, PSL Research University, CNRS, Sorbonne Universités, UPMC Univ. Paris 06, F-75014, Paris, France 
\and  Max-Planck-Institut f\"ur Radioastronomie, Auf dem H\"ugel 69, D-53121 Bonn, Germany  
}

   \date{Received June 22, 2015 /  Accepted November 30, 2015}

  \abstract
{We have used
the \emph{Herschel}-HIFI instrument    to observe the two nuclear spin symmetries of amidogen (NH$_2$) 
towards the high-mass star-forming regions
W31C (G10.6$-$0.4), W49N (G43.2$-$0.1), W51 (G49.5$-$0.4), and G34.3+0.1. The aim is  to investigate  the ratio of nuclear spin 
types, the ortho-to-para ratio (OPR) of NH$_2$   in 
the  translucent interstellar gas, where it is traced by the line-of-sight absorption, and
in the envelopes that surround the hot  cores. 
The HIFI instrument allows spectrally resolved observations of NH$_2$   that show a complicated pattern of hyperfine
structure components in all its rotational transitions.  
The  excited NH$_2$ transitions were used to construct radiative transfer models of the hot cores and surrounding envelopes 
to investigate the excitation and possible emission of  the ground-state rotational transitions of  
ortho-NH$_2$ \mbox{$N_{K_a, K_c}J$\,=\,1$_{1,1}\,3/2$\,--\,0$_{0,0}$\,1/2}  (953 GHz)
 and    para-NH$_2$ \mbox{2$_{1,2}\, 5/2$\,--\,1$_{0,1}\,3/2$} (1444 GHz) 
used in  the OPR calculations.  
 Our best estimate  of the average 
   OPR in the envelopes lie above the  high-temperature limit
   of three for    W49N, specifically 3.5 with formal errors of $\pm$0.1, but for W31C, W51, and G34.3+0.1 we find lower values of 2.5$\pm$0.1, 2.7$\pm$0.1, and 2.3$\pm$0.1, respectively. 
   Values this low are strictly forbidden in thermodynamical equilibrium since the OPR 
   is expected to increase above three at low temperatures. 
  In the translucent interstellar gas towards W31C, where the excitation effects are low,  we find similar values 
    between 2.2$\pm$0.2 and 2.9$\pm$0.2. In contrast, we find  an OPR  of  3.4$\pm$0.1  in the dense and cold filament connected to W51 and  also 
    two  lower limits of $\gtrsim 4.2$ 
    and $\gtrsim5.0$ in two other translucent gas components towards W31C and W49N. 
    At low temperatures ($T \lesssim 50$~K) the OPR of
H$_2$ is $<10^{-1}$, far lower than the terrestrial laboratory
normal  value of three. In this para-enriched H$_2$ gas, our
astrochemical models can reproduce  the variations of the observed   OPR, both below and above the  
thermodynamical equilibrium value, by considering nuclear-spin gas-phase chemistry. 
The models suggest  that values  below three arise in regions with temperatures   $\gtrsim20-25$~K, 
depending on time, and values above three at lower temperatures.
}

   \maketitle
%

\section{Introduction}
The amidogen (NH$_2$) radical is   
an important species in the first steps of reaction chains in the nitrogen chemical 
network, and it is 
closely related to the widely observed 
ammonia (NH$_3$) molecule. Observations of NH$_2$ can be used for testing the 
production pathways of nitrogen-bearing molecules. Despite its importance, few observations have  
been performed   of interstellar NH$_2,$  however, since  most   strong 
transitions fall at sub-mm wavelengths, which are difficult to observe using ground-based antennas.

Each rotational transition of NH$_2$ has   a complex fine and hyperfine   structure.
It  is a  light asymmetrical     rotor with 
two symmetry forms, which are caused by the different relative orientations of the hydrogen nuclear spins. 
The  symmetries are believed to  behave like two
distinct species: ortho-NH$_2$ (H spins parallel) and 
para-NH$_2$ (H spins anti-parallel). Its energy level diagram 
(on-line Fig.~\ref{NH2 energy level diagram})  
is similar to that  
of water, with the difference that NH$_2$ has exchanged  ortho and para symmetries, that is, the 
lowest level ($0_{0,0}$) is ortho. 
The electronic ground state is $^2B_1$  with a net 
unpaired electronic spin of 1/2, which splits each $N_{K_a, K_c}$ level  with 
$J = N \pm 1/2$. 
The nitrogen nuclei further split all levels, and the ortho-transitions also have 
additional splitting that is due to the net proton spin.   
We note that the spin degeneracy $(2I+1)$   makes the ortho stronger than the para lines.

 \begin{table*}[\!htb] 
\centering
\caption{Source sample  properties.}
\begin{tabular} {lccccc } 
 \hline\hline
     \noalign{\smallskip}
Source  & R.A.  & Dec. & Distance & $\varv_\mathrm{LSR}$\tablefootmark{a} & l.o.s.\tablefootmark{b}  \\ 
 &  ($J$2000)  &  ($J$2000)  & (kpc) & (km~s$^{-1}$) & (km~s$^{-1}$)\\
     \noalign{\smallskip}
     \hline
\noalign{\smallskip}  
W31C (G10.6$-$0.4)  & 18:10:28.7 & $-$19:55:50  & 4.95\tablefootmark{c} & $-$3.5 & 10--61 \\
W49N (G43.2$-$0.1) & 19:10:13.2&  +09:06:12  & 11.1\tablefootmark{d}  & +8.0 & 20--84 \\
W51  (G49.5$-$0.4) & 19:23:43.9&  +14:30:30.5 & 5.4\tablefootmark{e} & +57 & 1--45    \\
G34.3+0.15 & 18:53:18.7& +01:14:58 & 1.6\tablefootmark{f} & +58 & 8--45  \\
    \noalign{\smallskip} \noalign{\smallskip}
\hline 
\label{Table: sources}
\end{tabular}
\tablefoot{
\tablefoottext{a}{Source LSR velocity.} 
\tablefoottext{b}{LSR velocity range of foreground absorbing gas.} 
\tablefoottext{c}{\citet{2014ApJ...781..108S}.}  
\tablefoottext{d}{\citet{2013ApJ...775...79Z}.}
\tablefoottext{e}{\citet{2010ApJ...720.1055S}.}
\tablefoottext{f}{\citet{2011PASJ...63..513K}.} 
}
\end{table*}


 \begin{table*}[\!htb] 
\centering
\caption{Observed NH$_2$ transitions\tablefootmark{a} towards W31C, W49N, W51, and G34.3+0.1,
 resulting continuum intensities, and noise levels.
}
\begin{tabular} {lcccccccccccccc} 
 \hline\hline
     \noalign{\smallskip}
Sym.    & Freq.\tablefootmark{b}& Transition & $\theta_\mathrm{mb}$ & $\eta_\mathrm{mb}$      &    $E_\mathrm{l}$  
&    $E_\mathrm{u}$     &  \multicolumn{4}{c} {$T_\mathrm{C}$\tablefootmark{c}}   
 &\multicolumn{4}{c}{$1\sigma$/$T_\mathrm{C}$\tablefootmark{d}}  
\\    \noalign{\smallskip}
&&& &&& &W31C & W49N & W51 & G34.3  & W31C & W49N & W51 & G34.3  \\
& (GHz) &   $N_{K_a,K_c}\, J$ & ($\arcsec$)& & (K) &  (K)   & (K)& (K)& (K)& (K)  & &&    \\
     \noalign{\smallskip}
     \hline
\noalign{\smallskip}  

ortho   &       648.78423&       $ 2_{1,1} \, 3/2 - 2_{0,2}\, 3/2$ & 32 & 0.65 & 89.4   &    120      & 0.84  & 1.2 &  1.4&  1.2 &0.013& 0.008&   0.014 & 0.008  \\ %
 
ortho   &       907.43278&     $2_{0,2}\, 5/2 - 1_{1,1}\, 3/2$ &23& 0.63 &45.7     &    89.3     & 2.4 & 3.2 & 3.5& 3.0& 0.010 & 0.008& 0.010& 0.013 \\%

ortho   &       952.57835       &  $1_{1,1}\, 3/2 - 0_{0,0}\, 1/2$ & 22 &0.63&  0    &  45.7         & 2.7     & 3.7&  3.9 &  3.3& 0.017 & 0.018& 0.018 & 0.021  \\%

ortho   &       1\,012.43614    &  $4_{2,2}\, 9/2 - 4_{1,3}\, 9/2$ & 21&0.62 & 350       &   399  & 3.3&\ldots&\ldots&\ldots & 0.002  &\ldots&\ldots&\ldots    \\ %

para    &       1\,443.62839    &  $2_{1,2} 5/2 -  1_{0,1} 3/2$ & 15&0.59&  30.4       &    99.7      & 5.2  & 7.5 & 6.6 &  5.8 &0.012& 0.012 &0.013& 0.017  \\ %
   
    \noalign{\smallskip} \noalign{\smallskip}
\hline 
\label{Table: transitions}
\end{tabular}
\tablefoot{
\tablefoottext{a}{HIFI consists of seven  
mixer bands and two  double-sideband spectrometers. The 
649 transition was observed in the upper sideband of band 2a, the 907 and 953~GHz lines in the upper sideband of band 3b,  and para-NH$_2$ in the 
lower sideband of band 6a.  
} 
\tablefoottext{b}{The  frequencies were taken from the Cologne Database for Molecular Spectroscopy (CDMS) and refer to the strongest nuclear hyperfine component (hfs) in order to convert frequencies to
Doppler velocities relative to the local standard at rest $V_\mathrm{LSR}$. All hfs components can be found in Tables.~\ref{Table: 648 hfs transitions}-\ref{Table: 1444 hfs transitions}.} 
\tablefoottext{c}{The single-sideband (SSB) continuum intensity not corrected for beam efficiency.} 
\tablefoottext{d}{Rms noise  at a resolution of 0.5~km~s$^{-1}$  divided by $T_\mathrm{C}$.} 
}
\end{table*}

The first detection of interstellar  NH$_2$ was made by \citet{1993ApJ...416L..83V}  in 
absorption towards Sgr\,B2~(M). They detected the fundamental rotational transition \mbox{$N_{K_a, K_c} = 1_{1,0}-1_{0,1}$} of      
para-NH$_2$,  with  three fine-structure components at 461 to 469~GHz  and partially resolved hyperfine 
structure.
The Infrared Space Observatory (ISO) was later used 
by \citet{2000ApJ...534L.199C},  \citet{2004ApJ...600..214G}, and \citet{2007MNRAS.377.1122P} 
to observe spectrally unresolved 
absorption lines of both 
ortho- and para-NH$_2$  towards the same source. 
\citet{2004ApJ...600..214G} inferred an NH$_2$  ortho-to-para ratio (OPR) of  3$\pm$1 towards the
warm and low-density envelope around Sgr\,B2~(M) from the observation of
ortho-NH$_2$  $2_{2,0}-1_{1,1}$  and para-NH$_2$   $2_{2,1}-1_{1,0}$ rotationally excited far-IR
lines.
 
During its mission lifetime 2009-2013, the \emph{Herschel} Space Observatory
\citep{Pilbratt2010}, with the Heterodyne Instrument for the Far-Infrared \citep[HIFI;][]{Graauw2010}
offered unique capabilities for observations at very high sensitivity and high spectroscopic resolution  
of the fundamental 
rotational transitions of light hydrides at THz frequencies (0.48-1.25~THz and 1.41-1.91~THz). 
The PRISMAS key programme (PRobing InterStellar Molecules with Absorption line Studies)   targeted 
absorption lines along the sight-lines towards eight bright sub-millimetre-wave continuum sources using \emph{Herschel}-HIFI: 
W31C (G10.6$-$0.4), W49N (G43.2$-$0.1), W51 (G49.5$-$0.4), G34.3+0.1, DR21(OH), SgrA (+50~km~s$^{-1}$ cloud), W28A (G005.9$-$0.4), and W33A. 
With this technique,  the interstellar gas along the sight-lines can be seen in absorption, simultaneously 
with the hot cores that are detected through emission and absorption. 
 
The first results and analysis of absorption lines of nitrogen hydrides (NH, NH$_2$, and NH$_3$) along the sight-lines  towards the massive star-forming 
regions W31C  and W49N  
were presented in 
\citet[paper I,][]{2010A&A...521L..45P} and \citet[paper II,][]{2012A&A...543A.145P}. 
Similar average abundances with respect to 
the total amount of hydrogen 
estimated along the whole line of sight towards W31C were found in paper~I for all three species,     
$\sim6\times 10^{-9}$,  $3\times 10^{-9}$, and $3\times 10^{-9}$ for NH, NH$_2$, and NH$_3$, 
respectively. These abundances were, however, estimated from the ortho NH$_2$ and NH$_3$ symmetries 
using the high-temperature OPR limits of 3 and 1, respectively.    
In paper~II,  absorptions along the sight-lines towards W31C and W49N were decomposed  
into different velocity components. Column densities and abundances with respect to molecular
hydrogen   were estimated in each velocity
component, and a linear correlation  of  ortho-NH$_2$ and  ortho-NH$_3$ was found with a ratio of $\sim1.5$. 
We  also found surprisingly low OPR ammonia  values, $\sim0.5-0.7$,  below  the strict high-temperature limit 
of unity. 
Values as low as this are strictly forbidden in thermodynamical equilibrium, since the OPR is expected to be unity 
at temperatures significantly higher than 22~K, the energy difference between ortho- and para-NH3, and 
increase above unity at low temperatures.  
As suggested by \citet{2013ApJ...770L...2F} and further developed by \citet{2014A&A...562A..83L},  
a low OPR is a natural consequence of 
chemical gas-phase reaction pathways and is consistent with nuclear spin selection rules in a para-enriched 
H$_2$  gas that drive the OPR of nitrogen hydrides to values lower than the statistical limits. 
Their prediction  for NH$_2$,
which is  directly chemically related to NH$_3$, is an OPR of
$\sim$2.0-2.8 depending on physical conditions and initial abundances. 
This is in contrast to the expected increase above 
the high-temperature limit of 3  at low temperatures. 

Our aim in this paper is to investigate the behaviour of the 
OPR of NH$_2$ towards four of the PRISMAS sources: W31C, W49N, G34.3+0.1 (hereafter G34.3), and W51. To do this, we  
use PRISMAS data and observations of additional higher excitation transitions of NH$_2$ 
from our   programme OT1$\_$cpersson$\_$1 \emph{Investigation of the nitrogen chemistry in diffuse and dense interstellar gas}.


 \section{Observations and data reduction}
The observed high-mass star-forming regions and their properties are
listed in Table~\ref{Table: sources}, and the \emph{Herschel} observational identifications (OBSIDs) can be found 
in Table~\ref{Table: obsid} (on-line material).   
Towards all four sources we have observed 
three \mbox{ortho-NH$_2$}   and 
one para-NH$_2$  line (Table~\ref{Table: transitions}). 
For the spectroscopy we consulted the Cologne Database for Molecular 
Spectroscopy\footnote{{\tt http:www.cdms.de}} \citep[CDMS;][]{2001A&A...370L..49M, 2005JMoSt.742..215M}  
based on spectroscopy   by \citet{1999JMoSp.195..177M}. 
A fifth line, the ortho 1012~GHz line, was also detected towards W31C  
while searching for NH$^+$    \citep{2012A&A...543A.145P}.  
The on-line Tables.~\ref{Table: 648 hfs transitions}-\ref{Table: 1444 hfs transitions} list 
the hyperfine structure  (hfs) components of all observed transitions.  
The observations of the  $1_{1,1} - 0_{0,0} $ 953~GHz transition towards 
W31C and W49N have already been presented in papers~I and II.

We used the
dual beam switch mode and the wide band spectrometer (WBS) with a bandwidth of 
4~GHz for the lower bands, and 2.5~GHz for band 6, with an effective spectral  
resolution of 1.1~MHz. 
The corresponding  velocity resolution is  0.51,  0.37, and 0.23~km~s$^{-1}$ at 649, 
907, and 1444 GHz, respectively. 
Two
orthogonal polarisations  
were used in all observations. 
The ortho lines were observed with three different frequency settings of the local 
oscillator (LO), and the para line with five settings,  
corresponding to a change of approximately 15\,km\,s$^{-1}$~to
determine the sideband origin
of the lines, since HIFI uses double-sideband (DSB) receivers. No contamination from the image sideband was detected except for 
several lines blended with the ortho 649~GHz  line  in all settings towards W49N, W51, and G34, which we were unable to remove. We use these observations as upper limits of the true 649~GHz line emission in our modelling 
in Sect.~\ref{ali}.    
The ortho 953~GHz absorption along the line-of-sight gas towards W49N is contaminated by  an NO emission line  from the source in the same sideband,  
which was removed before the analysis (described in paper~II).

The recommended values for half-power beam width of the telescope and the main beam-efficiencies are  
listed in Table~\ref{Table: transitions} and taken from the HIFI release note 2014\footnote{{\tt http://herschel.esac.esa.int/twiki/pub/Public/Hifi-\\CalibrationWeb/HifiBeamReleaseNote\_Sep2014.pdf}}.  
The in-flight performance is  
described by \citet{2012A&A...537A..17R} and the calibration uncertainties are $\lesssim$9\% for band 3 and $\lesssim$11\% for band~6.

The data were processed using the standard \emph{Herschel} Interactive Processing Environment  
  \citep[HIPE,\footnote{HIPE is a joint development by the \emph{Herschel} Science Ground
Segment Consortium, consisting of ESA, the NASA \emph{Herschel} Science Centre, and the HIFI, PACS and
SPIRE consortia.}][]{2010ASPC..434..139O},
version 12.1,   up to level 2 providing fully calibrated DSB
spectra
on the  $T_\mathrm{A}^*$ antenna temperature intensity  
scale where the lines are calibrated to single sideband (SSB) and the continuum to DSB.  
The FITS  files were then exported to  the spectral line   software package  {\tt xs}\footnote{Developed by 
Per Bergman at Onsala Space Observatory, Sweden; {\tt http://www.chalmers.se/rss/oso-en/observations/data-\\reduction-software}} 
, which was used in the subsequent data reduction. 
All LO-tunings and both polarisations were included in the averaged noise-weighted   spectra, except for a few 
settings of para-NH$_2$  which suffered from severe spikes (details in Table~\ref{Table: obsid}). 
Absorption from HCl$^+$ has been detected at  lower
velocities ($\sim40-210$~km~s$^{-1}$)  close to  our observed para-NH$_2$ line   \citep{2012ApJ...751L..37D}. 
This explains why the base
line seems to present undulations at these frequencies. 
The averaged spectra were finally convolved from the original 0.5~MHz channel separation to   the effective spectral resolution of 1.1~MHz. 
In order to derive the OPR, we resampled the para spectra to the  velocity resolution of the ortho spectra, 
which allows   comparison of column densities in individual velocity bins.

 The resulting noise and SSB continuum levels (not corrected for beam efficiency), $T_\mathrm{C}$,  as measured in line-free regions in the spectra, 
 are found in Table~\ref{Table: transitions}. We note 
that since HIFI uses DSB receivers, the observed continuum has to be 
divided by two to obtain the SSB continuum.  
The sideband gain ratio is assumed to be unity throughout this paper. 
Unless otherwise specified, all spectra in this paper are shown in a $T_\mathrm{A}^*$ SSB scale.


 \section{Results and modelling} \label{section: results}

\begin{figure} 
\resizebox{\hsize}{!}{
\includegraphics{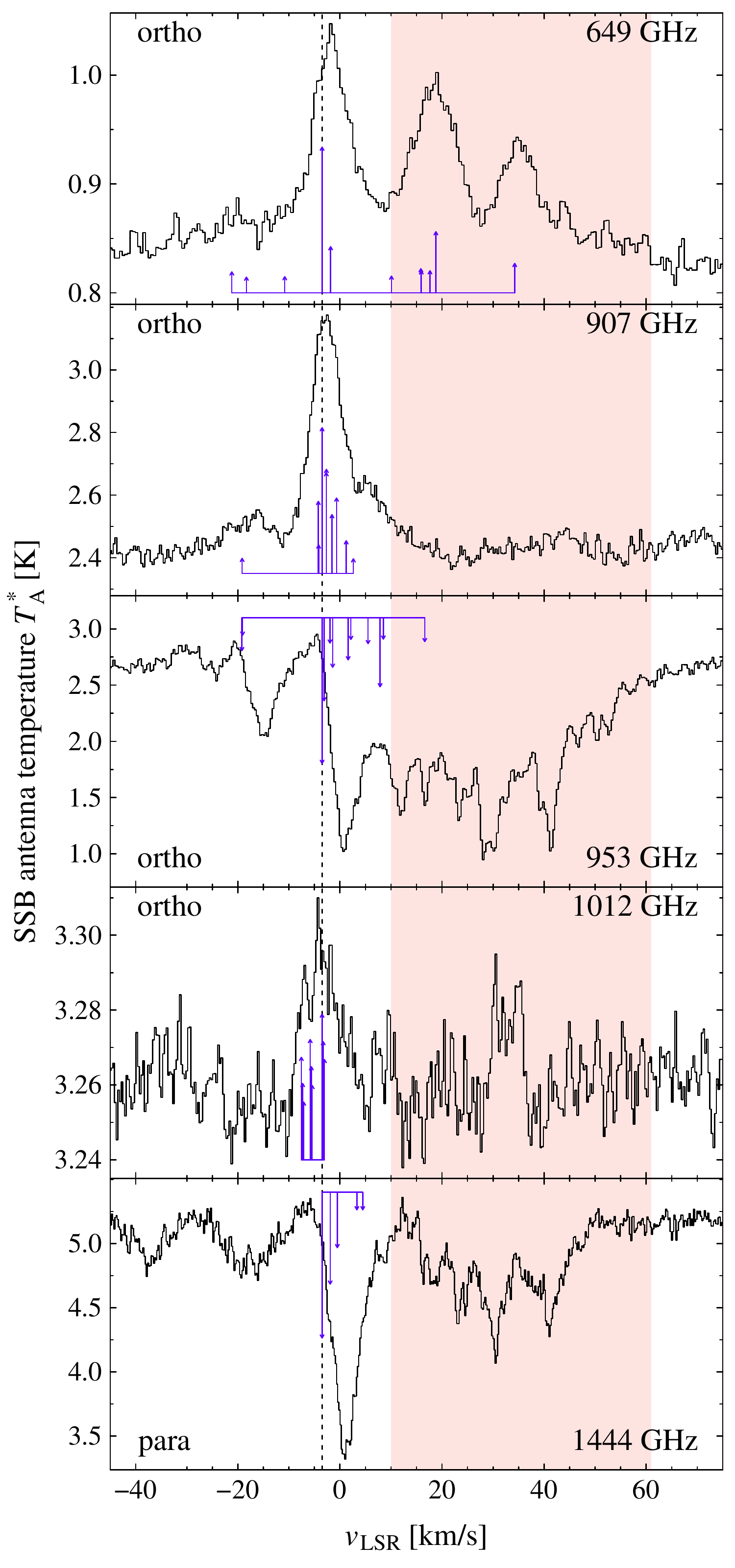}}  
\caption{ 
Single-sideband   spectra of   NH$_2$   towards W31C. The blue arrows mark  the relative positions 
and intensities of the strongest hyperfine structure components (>10\% of the main hfs). The dotted line shows the $V_\mathrm{LSR}$
of the source and the red box the  $V_\mathrm{LSR}$ range for   the line-of-sight interstellar gas.
}
\label{Fig: W31C all lines}
\end{figure}

\begin{figure} 
\resizebox{\hsize}{!}{
\includegraphics{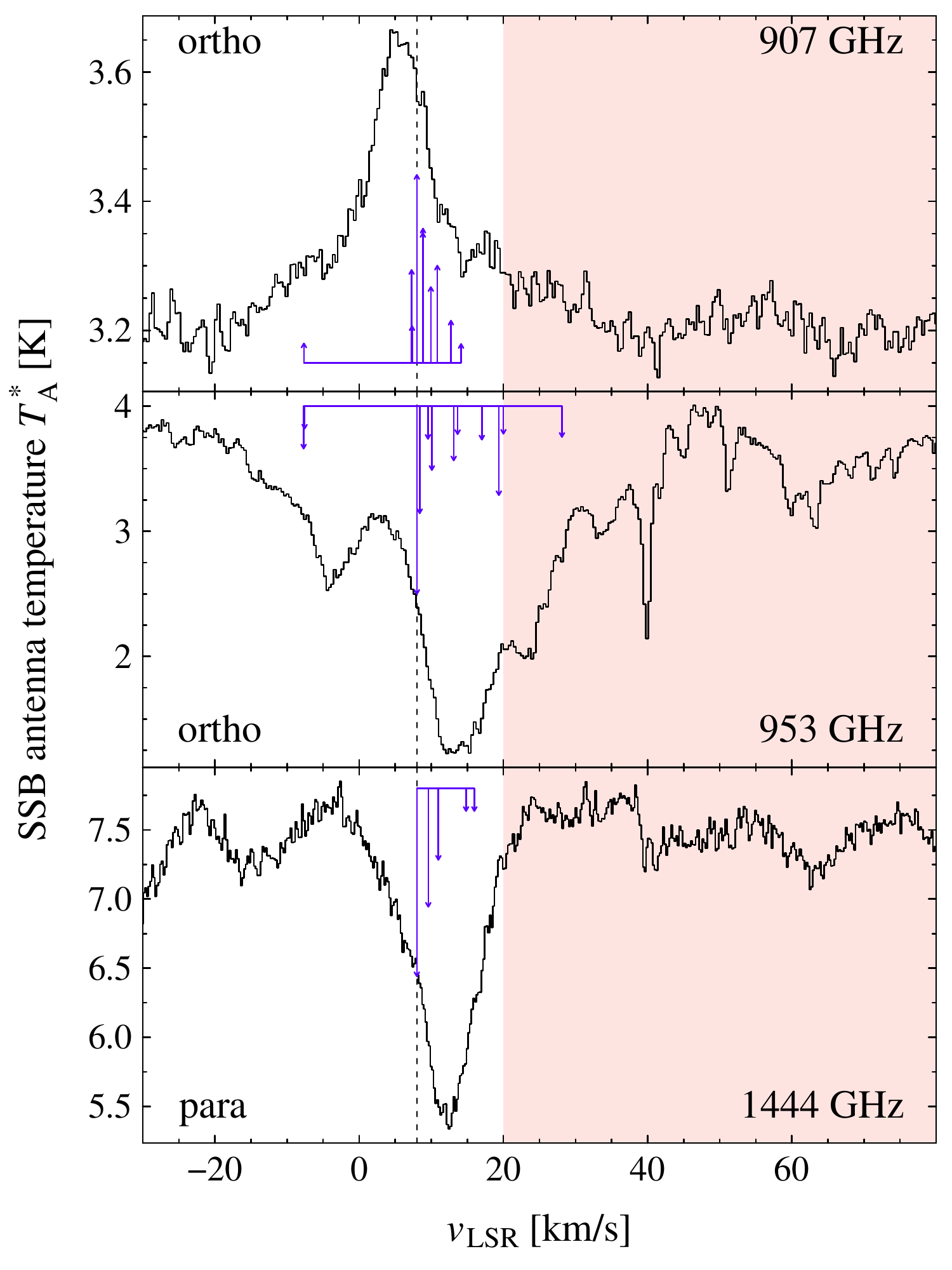}}  
\caption{ 
W49N. Notation as in Fig.~\ref{Fig: W31C all lines}. 
}
\label{Fig: W49N all lines}
\end{figure}

\begin{figure} 
\resizebox{\hsize}{!}{
\includegraphics[angle=0]{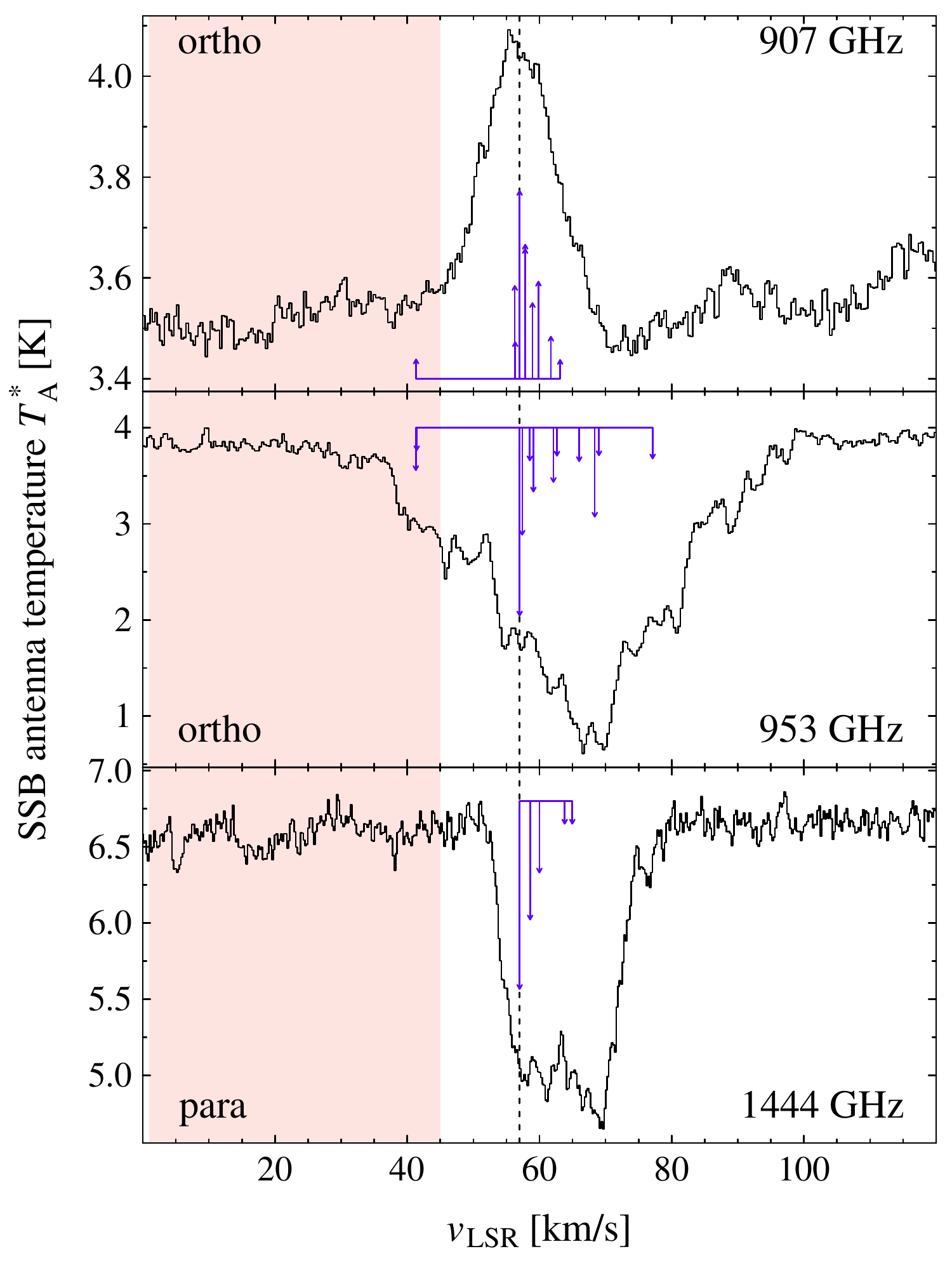}}  
\caption{ 
W51. Notation as in Fig.~\ref{Fig: W31C all lines}. 
}
\label{Fig: W51 all lines}
\end{figure}

\begin{figure} 
\resizebox{\hsize}{!}{
\includegraphics{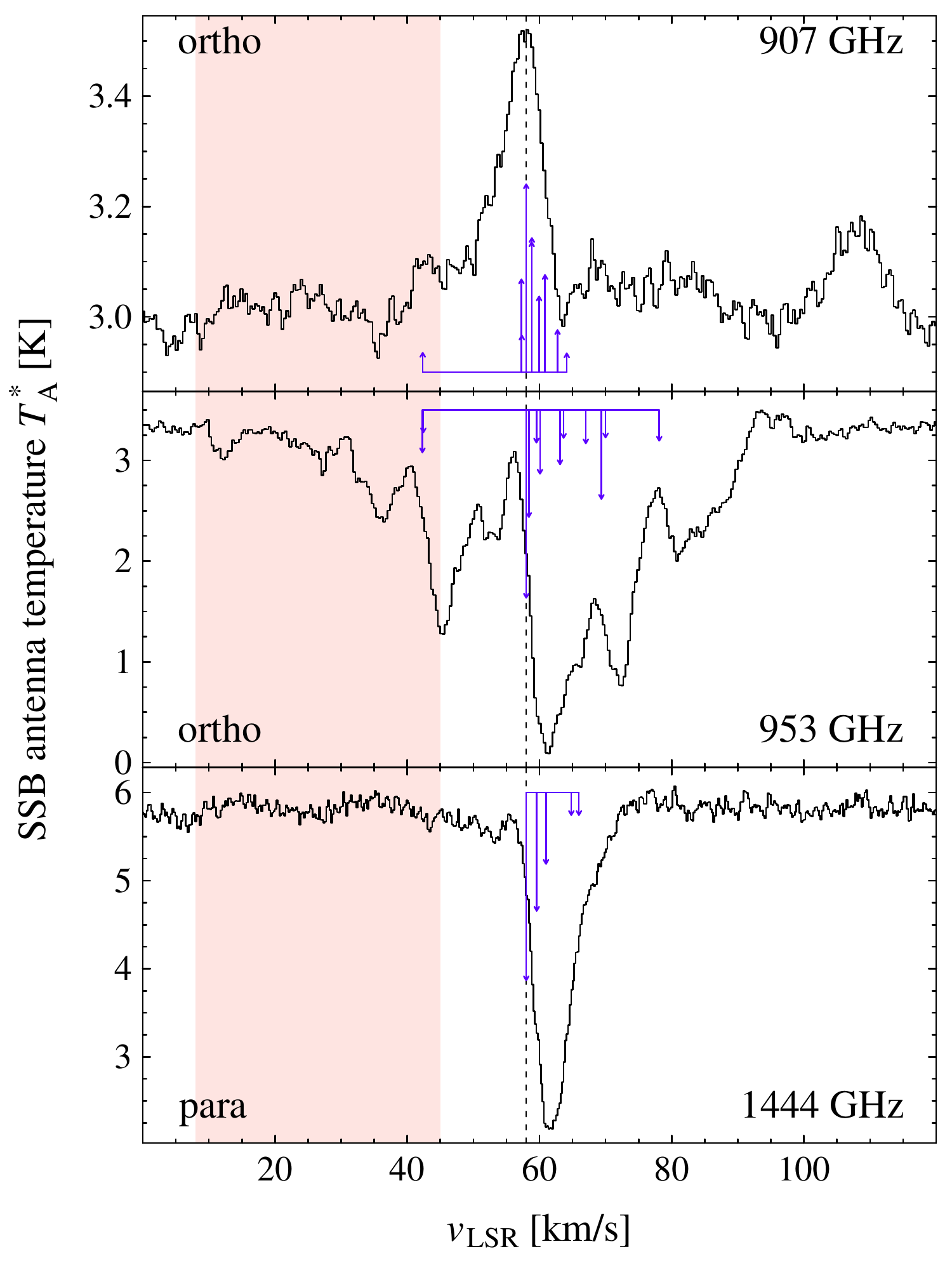}}  
\caption{ 
G34.3+0.1. Notation as in Fig.~\ref{Fig: W31C all lines}. 
}
\label{Fig: G34 all lines}
\end{figure}

The resulting spectra are shown in 
Figs.~\ref{Fig: W31C all lines}--\ref{Fig: G34 all lines}. The different line shapes reflect not only the different hyperfine
structures (marked with blue arrows in the figures), but also the fact that they trace different regions within the beam. 
The more highly excited ortho-lines  are detected in emission from the hot cores, while the 
lowest ortho- and 
para-NH$_2$ lines at 953 and 1\,444~GHz   show absorption lines   close to the source 
systemic velocities from the envelopes surrounding the hot cores, and also 
in a wide range of velocities tracing the translucent line-of-sight gas towards W31C. 
The sight-line towards W49N  shows   
absorption components in the ortho line, but no detection of the   para line  within the noise limits. 
Neither ortho nor para absorptions are detected   
towards W51 and G34.3+0.1   at 
\mbox{$V_\mathrm{LSR} \lesssim 45$~km~s$^{-1}$} and \mbox{$ \lesssim 50$~km~s$^{-1}$}, respectively. 
This is in 
contrast to many other species,  for example, CH \citep{2010A&A...521L..16G}, 
H$_2$O   \citep[][]{2013ApJ...762...11F} and HF \citep{2010A&A...521L..12S}, which trace lower density interstellar gas than NH$_2$
\citep[see e.g. chemical models for the nitrogen hydrides in different physical conditions in][]{2014A&A...567A.130P}.  
We did, however, find  redshifted absorption in W51, 
at \mbox{$V_\mathrm{LSR} \sim68$~km~s$^{-1}$}, tracing   
a dense clump in a filament interacting with W51, also  detected in C$_3$, HDO and NH$_3$   
 \citep{2014A&A...566A..61M}.   
The absence of emission  from any of the rotational  NH$_3$ transitions in this component 
  suggests  a   cold core.

In summary, our approach to  estimating  the OPR is to 
compare the absorption of the lowest ortho and para lines at 953~GHz and 1444~GHz, respectively. 
We calculate the opacities in each velocity bin, 
and then convert this opacity ratio into a column density ratio. 
In this way, we  
find the OPR in different types of physical conditions traced by the absorption lines: 
the   molecular envelopes surrounding the hot cores,  
the line-of-sight translucent gas, and one dense and cold
core (the filament interacting with W51). 
 For the hot cores themselves it is not possible to reliably estimate the OPR, however, since we only
 have one para line that mainly traces lower excitation gas in the envelopes.

However,  before the above procedure was performed we addressed  three problems: 
\emph{(i)} The complex hyperfine structure of the different transitions that  prevents 
a  direct comparison of the spectra.   
This was solved by deconvolving the observed lines  with respective   hfs patterns. 
(The line blending prevents  an independent calculation of the opacities from 
the observed hyperfine structure components.) 
\emph{(ii)} Emission from the hot cores that    may be present at a low level in the  953 and 1\,444~GHz lines, although
  hidden by the strong absorption of the foreground  molecular envelope.  
By assuming that the emission from the  
hot core is separated from the absorption of the foreground molecular envelope, we used the more highly excited transitions  to
construct a model of the hot core emission. The  modelled emission line profiles of the 953 and 1\,444~GHz lines 
were then  removed from the observed spectra. 
\emph{(iii)} The excitation in the molecular envelopes themselves   affects the  absorption depths, and 
the derived opacities will be underestimated. Hence   the excitation temperature   cannot be neglected in these regions.   
This is, however, most likely not a major 
problem in the translucent interstellar gas, where the  
excitation is most likely low or completely negligible.  

Below we describe the details of  our analysis and modelling.

\subsection{ALI modelling: Hot core emission and excitation in the envelopes}\label{ali}
To   find the underlying emission
line structure of the 953~GHz ortho and 1\,444~GHz para lines in  the hot cores
and to examine the excitation in the envelopes, 
we used a non-LTE, one-dimensional, radiative transfer model based on the accelerated lambda iteration (ALI) scheme \citep{1991A&A...245..171R} to 
solve the radiative transfer  in a spherically symmetric model cloud.  
ALI has been tested and used, for example by \citet{2008A&A...479..779M} and \citet{2014ApJ...788L..32W}. 
We also used this code in our   current  modelling of our 
\emph{Herschel}-HIFI observations of seven  rotational ammonia transitions in G34.3  \citep{Hajigholi}, 
and W31C, W49N and W51 (Persson et al. in prep).  
Collision rates for neutral impact on NH$_2$ were estimated based upon an assumed quenching rate
coefficient of $5\times 10^{-11}$~cm$^3$~s$^{-1}$ and state-specific downward rates for radiatively
allowed transitions that scale in proportion to radiative line strengths. 

For our NH$_2$ observations, we constructed a simple homogeneous spherical clump model  for the hot core emission in all sources. 
This model was fitted to the excited emission lines that show no or very little evidence
of self-absorption. 
In a second model, fitted to all lines seen in both emission and absorption, we added an outer envelope with   lower density and temperature in an attempt to also 
reproduce the major features of  the  absorption, in order to derive an estimate of the excitation temperature in the outer envelopes.  
 The best fits were found by varying the 
temperature and density   until the modelled  649~GHz line  matched the observed one.  
The relative strengths of the 649 and 907~GHz lines 
are  sensitive to the source size and only weakly 
dependent on dust temperature; the 649~GHz line becomes weaker with respect to the
907~GHz line with increasing size.  
We then searched for a 
combination that produced the best agreement with the continuum levels for all transitions.  
The large number of hfs components makes the modelling very time-consuming,
however. 
We acknowledge that we therefore have not fully explored the large   parameter space, and 
 other solutions  that fit our observed data are thus certainly possible. For our purpose, however, other solutions 
 only have a minor impact on  the modelled emission line profiles for the 953~GHz   line,   
shown in Figs.~\ref{Fig: W31C ali emission}-\ref{Fig: G34 ali emission}, which were removed from the 
observed spectra together with the continuum in the normalisation (Sect.~\ref{OPR results}). 
The modelled emission from  the hot cores in the para line is all below the noise levels 
and was therefore not removed from the observed spectra.

\begin{figure} 
  \centering
 \resizebox{\hsize}{!}{
\includegraphics{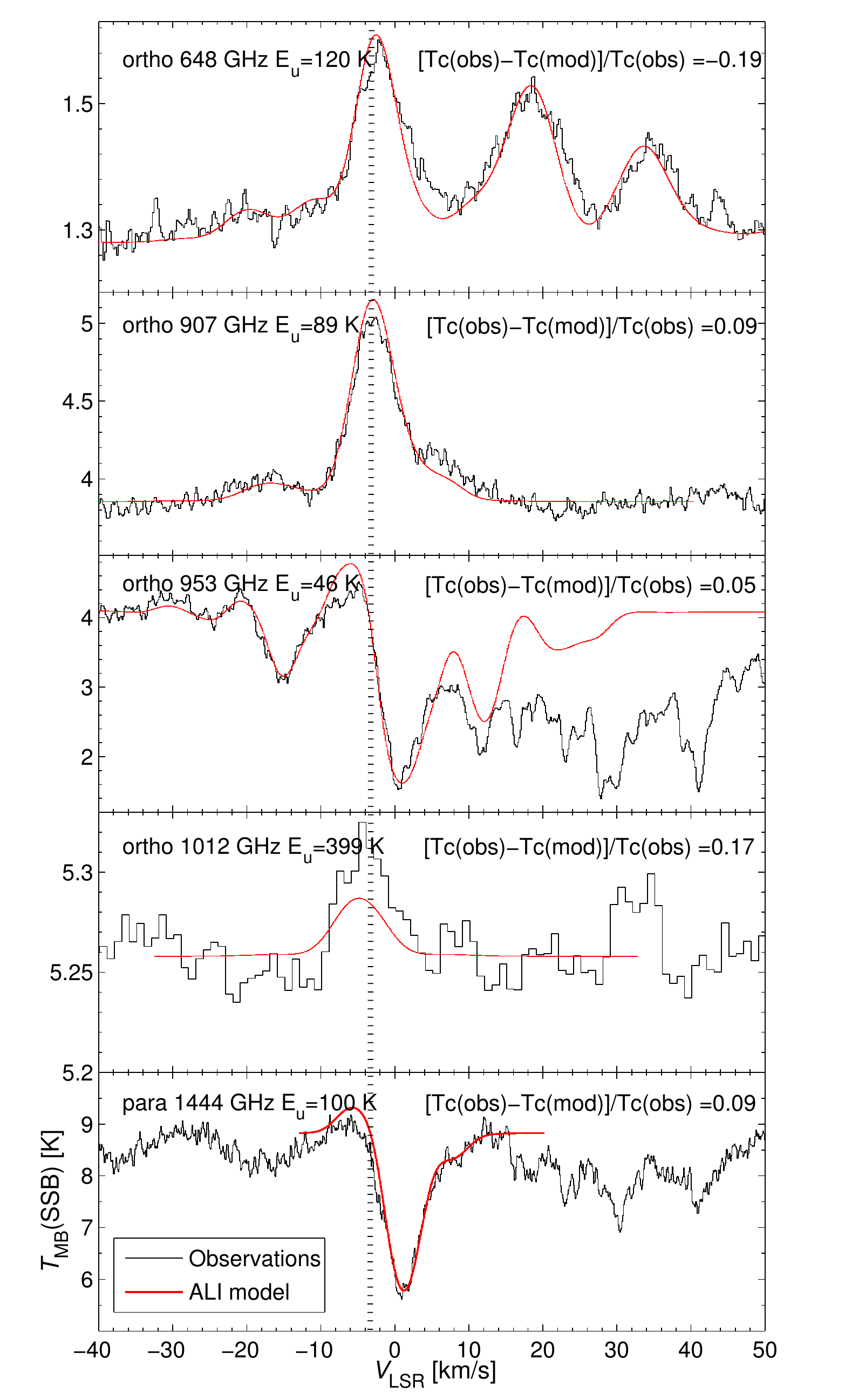} }
\caption{W31C: A two-component ALI model   of the hot core emission at \mbox{$V_\mathrm{LSR}=$-$3.5$~km~s$^{-1}$} 
(marked with a vertical dashed line),  
and foreground envelope absorption at \mbox{$V_\mathrm{LSR}=0$~km~s$^{-1}$}.  The absorptions from interstellar
gas along the sight-lines at \mbox{$V_\mathrm{LSR}=11-50$~km~s$^{-1}$}  
are not modelled. 
The relative difference of the modelled  and observed continuum
at the frequency of each line is given in the legend.
The resulting ortho-to-para ratio for this model  is 2.6 in the envelope.
} 
 \label{Fig: W31C ali 2 shell model}
\end{figure}

Our attempt to use a simple two-component model to reproduce the full line profiles 
  was successful towards W31C, 
where our best-fit model to all lines is shown in 
Fig.~\ref{Fig: W31C ali 2 shell model}. In this model, the ortho abundances in the hot core and envelope are $5\times10^{-10}$ and $2\times10^{-9}$, respectively, and the OPR in the envelope is 2.6.  
The excitation temperatures for the ortho and para line are 12.8 and 17.4~K, respectively.

The ortho 953~GHz line  was  much more difficult to model in the  remaining sources --   a more
complex density, temperature, and velocity distributions are most likely required.  
Examples  of (unsuccessful) models of both the emission and absorption towards W49N, W51 and G34.3 are shown 
in  the on-line  Figs.~\ref{Fig: W49N ali 2 shell model}-\ref{Fig: G34 ali 2 shell model}. 
The    ALI input parameters for all sources are listed in the on-line Table~\ref{table: ali parameters}, 
including the resulting excitation temperatures  used in Sect.~\ref{Subsect: excitation} to correct the opacities. 
Since we  did not succeed to model the depth of the ortho absorption, we used the derived excitation temperatures
as upper limits.


\subsection{Excitation effect on the opacity}\label{Subsect: excitation}

The optical depths per unit velocity interval, $\tau_\nu$, are derived from  
\begin{equation}\label{RTE}
T_\mathrm{MB} = T_\mathrm{C, MB}\,e^{-\tau_\nu} + J(T_\mathrm{ex})\,(1-e^{-\tau_\nu}) \  ,
\end{equation} 
where  $T_\mathrm{MB}$ and $T_\mathrm{C,MB}$ are the observed antenna and continuum temperature, respectively, 
corrected for the beam efficiency,   and
\begin{equation}\label{antenna temp}
J(T_\mathrm{ex}) = \frac{h\nu_\mathrm{ul}}{k} \frac{1}{\exp(h\nu_\mathrm{ul}/(k\,T_\mathrm{ex}))-1} \  ,
\end{equation} 
where $\nu_\mathrm{ul}$ is the frequency 
of the transition, and
$T_\mathrm{ex}$   the excitation temperature. We here assume that the  foreground absorbing material 
completely fills the beam and covers the continuum. This assumption is supported by the ALI modelling in Sect.~\ref{ali}. 
 
Solving Eq.~\ref{RTE} for the opacity, we find 
\begin{equation} \label{opacity}
\tau_\nu = -\ln \left( \frac{T_\mathrm{MB} - J(T_\mathrm{ex})}{T_\mathrm{C, MB} - J(T_\mathrm{ex})} \right ) \approx - \ln \left(\frac{T_\mathrm{MB}}{T_\mathrm{C,MB}}\right )  \ ,
\end{equation}
where the last approximation is only valid when 
 \mbox{$T_\mathrm{ex}\!\ll\! h\nu_\mathrm{ul}/k = 46$} and 69~K for the ortho and para line, respectively. 
 At low temperatures, the main influence of the excitation temperature is on the ortho   line, since  
the quantity $J(T_\mathrm{ex})$ is a sensitive function of frequency  and the excitation temperature, as shown in 
the on-line Fig.~\ref{Fig: JTex vs Tex}. Taking
into account the non-thermalised populations and thus different excitation temperatures    for the ortho and 
para species, we find that when using the excitation temperatures derived
with the ALI code (listed in Table~\ref{table: ali parameters}), where $T_\mathrm{ex}$ for ortho is lower than for para, 
the differences in $J(T_\mathrm{ex})$ decrease to $\sim1-10\, \%$. 
However, the stronger continuum at the frequency of the para line reduces the effect of excitation on this line. Neglecting the
excitation   gives  opacities of 0.6 and 0.27 of the main hyperfine components of the ortho 
and para lines, respectively, in the molecular envelope of W31C. 
Including the  $J(T_\mathrm{ex})$ correction, the opacities in the respective lines increase to 0.9 and 0.31. 
However, for the other three sources, where the excitation temperature is higher, the effect on the OPR from the para line increases due 
to the rapid increase of $J(T_\mathrm{ex})$.


\subsection{From opacities to column densities}\label{radex}
Since we were unable to model all sources with ALI,
we used  
the non-equilibrium homogeneous radiative transfer code 
{{\tt RADEX}}\footnote{\tt{http://www.sron.rug.nl/$\sim$vdtak/radex/index.shtml}}   \citep{2007A&A...468..627V} 
  to convert   opacities into column densities,  
taking into account both observed and unobserved   levels.  We used the same collision rates as for the ALI modelling. 
The principal input parameters are the molecular hydrogen density and kinetic temperature $T_\mathrm{K}$. The column 
density was varied until the integrated opacity was unity. 
The conversion factors   were estimated 
in the ranges  \mbox{$n(\mathrm{H}_2) = 500-3\,000$~cm$^{-3}$} and kinetic temperatures \mbox{$T_\mathrm{K} = 30-100$~K} for the  translucent interstellar gas, 
and \mbox{$n(\mathrm{H}_2) = 10^4-10^6$~cm$^{-3}$} and \mbox{$T_\mathrm{K} = 15-70$~K} for the source molecular envelopes and the dense, cold
clump in W51. 
The results are not very sensitive to changes in density  or temperature because of the high critical 
densities of the NH$_2$ transitions ($n_\mathrm{crit} =10^8-10^9$~cm$^{-3}$). For the line-of-sight gas, we used the average 
Galactic background radiation in the solar neighbourhood plus the cosmic microwave background 
radiation as background radiation field, and for the source molecular clouds we included the 
respective observed spectral energy distribution. 
In translucent cloud conditions, the 
integrated opacity is 1.0~km~s$^{-1}$ for  the 953~GHz and 1\,444~GHz transitions 
 when 
\mbox{$N($ortho-$\mathrm{NH_2}) = 3.6\times$10$^{12}$}~cm$^{-2}$, and \mbox{$N($para-$\mathrm{NH_2}) = 8.1\times$10$^{12}$~cm$^{-2}$}, respectively.  
The ratio of the conversion factors is 2.25, which corresponds to a factor of 6.8 for an OPR of  three. 
The ratio is similar  to within $\lesssim20~\%$ for  all our investigated physical conditions.

\subsection{Ortho-to-para results}\label{OPR results}

\begin{figure} 
  \resizebox{\hsize}{!}{
\includegraphics{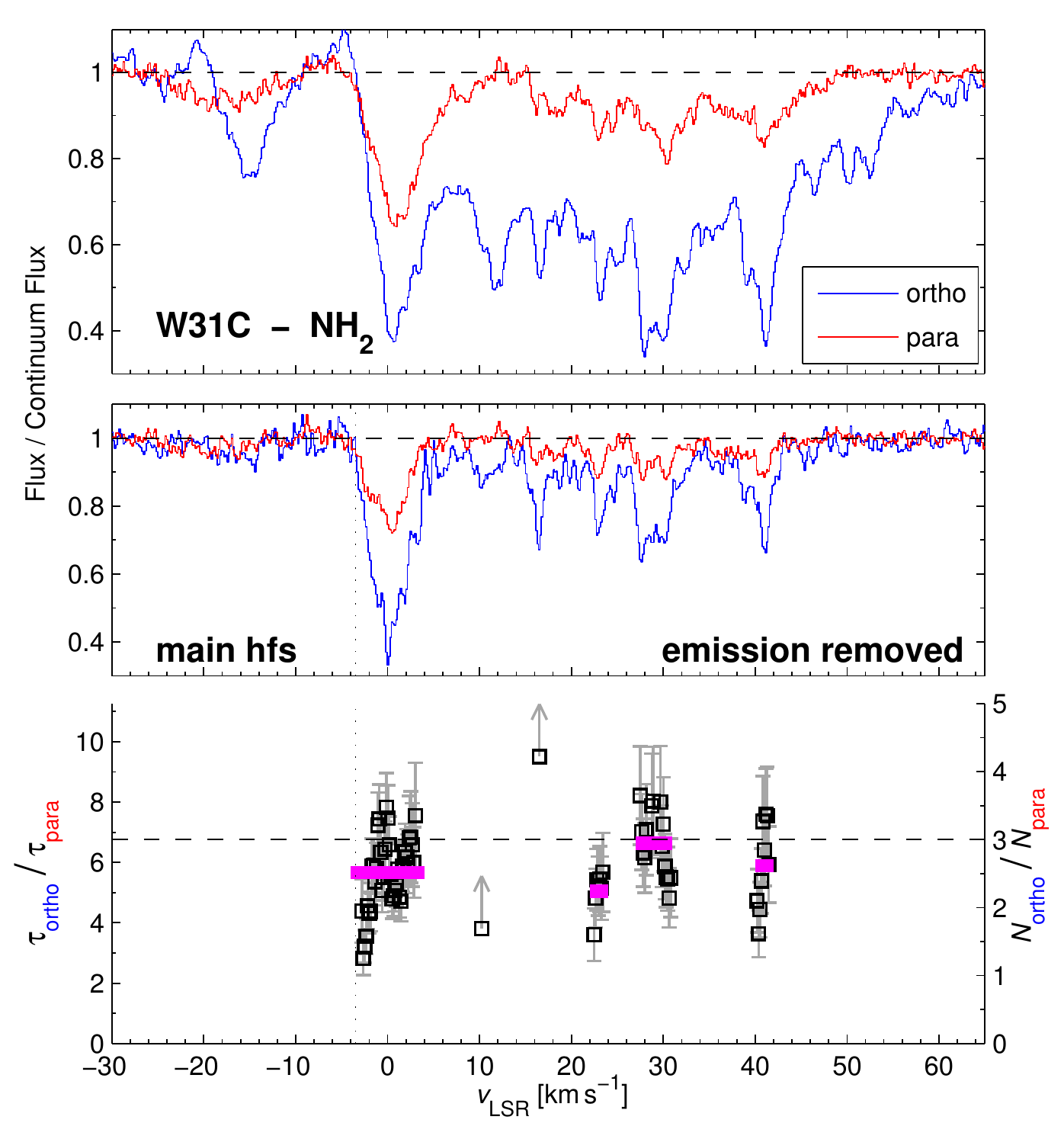}}  
\caption{W31C:  \emph{(Upper)} Normalised   spectra of the 
ortho   953~GHz and   
para 1\,444~GHz lines.  
\emph{(Middle)} Deconvolved      spectra where the strongest hfs component is plotted for both transitions. 
The 
hot core $V_\mathrm{LSR}$ is marked with a dotted vertical line. 
\emph{(Lower)} 
The optical depth and column density ratios of the convolved spectra as  functions of $V_\mathrm{LSR}$ for absorptions larger than 5\,$\sigma$.  The horizontal dashed line marks the high-temperature OPR limit of three.  
(Details are found in Sect.~\ref{OPR results}.)
} 
 \label{Fig: W31C Orto and para NH2}
\end{figure}

\begin{figure} 
  \resizebox{\hsize}{!}{
\includegraphics{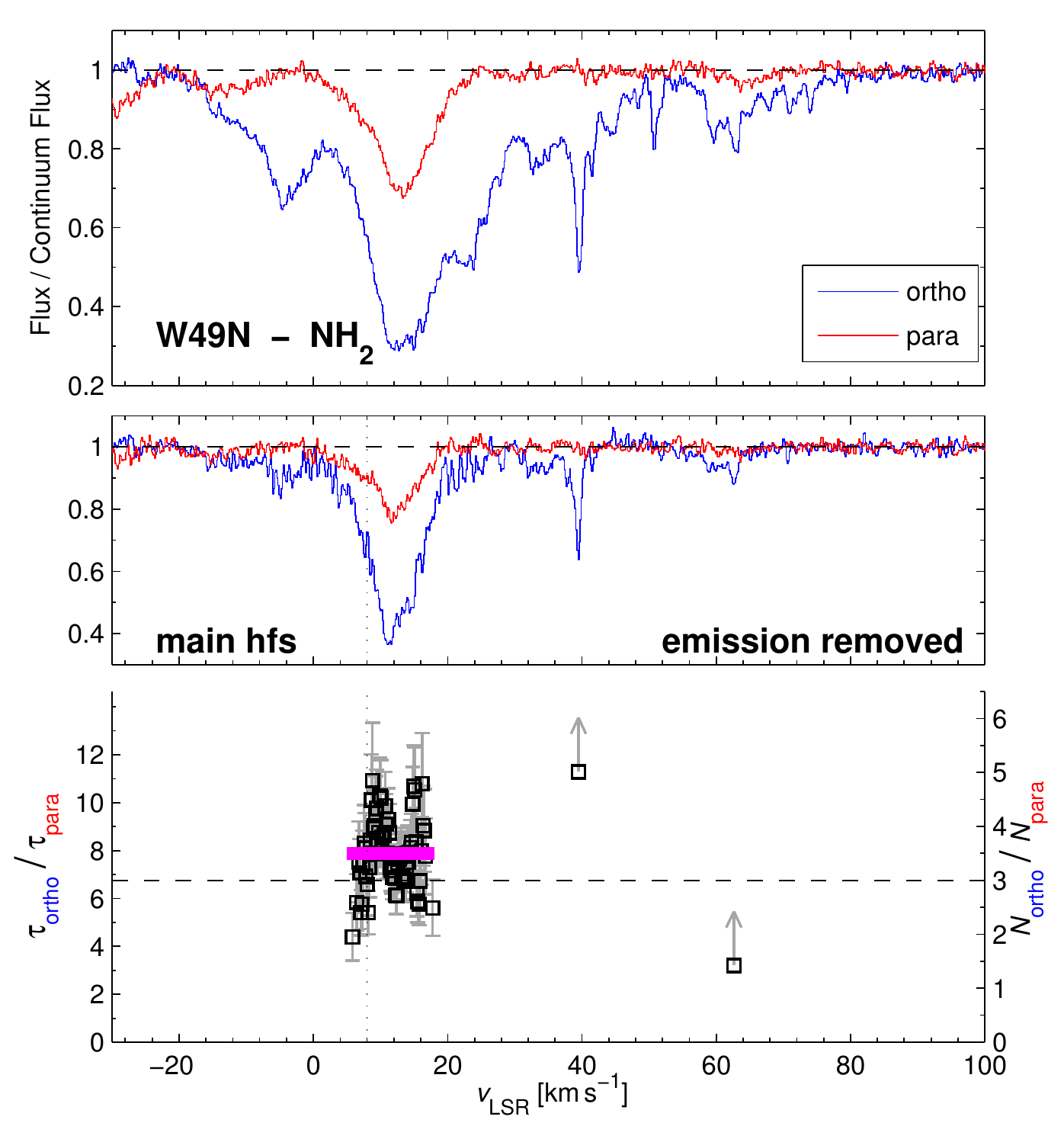}}  
\caption{W49N: Notation as in Fig.~\ref{Fig: W31C Orto and para NH2}. 
} 
 \label{Fig: W49N Orto and para NH2}
\end{figure}

The upper panels in 
Figs.~\ref{Fig: W31C Orto and para NH2}--\ref{Fig: G34 Orto and para NH2} 
show   the  ortho  953~GHz and    para 1\,444~GHz lines, where the intensities have been 
normalised to the continuum in a single sideband as $T_\mathrm{A}^*/T_\mathrm{C}$. 
The middle panels show the spectra after three corrective steps have
been applied:
\begin{itemize}
\item  Emissions from the central sources were removed by adding modelled
       emission profiles  to the continuum levels
       that in turn were used for the normalisation (emission is treated as a
       variation in the background continuum).
\item  A deconvolution algorithm was applied to provide spectra
       that only contain the main hfs component. The assumption used is that
       in the optical depth domain, our spectra can be treated as a set of single-line
       velocity components that have been convolved with an hfs structure
       according to the expected LTE relative optical depths (on-line \mbox{Tables~\ref{Table: 648 hfs transitions}--\ref{Table: 1444 hfs transitions}}). 
\item     We include correction for the excitation  
       by applying our  best estimate  of $T_{\rm ex}$\ in a velocity range around
       the source velocity, meaning that in these ranges the normalisation is made
       through 
       \mbox{$(T_\mathrm{MB} - J(T_\mathrm{ex}))/(T_\mathrm{C, MB} - J(T_\mathrm{ex}))$}. 
       No corrections were made for the line-of-sight components. 
\end{itemize}
In the lower panels, the left hand y-axis  shows the   optical depth ratios for the strongest hfs components. 
The corresponding column density ratios, summed over all hfs components 
(a correction factor of 4.5/2.25),     are given by the right hand  y-axis.  
To convert from opacity ratios into column density ratios, we used the conversion 
factors obtained with {\tt RADEX}.  
We only included data
points where the absorption depths of both ortho and para exceeded five times
the thermal rms, as measured in line-free regions of the baseline. In a few
cases we show lower limits, where only the ortho line satisfies this condition.
Assuming that the OPR is constant across the line profile in each 
velocity component,  we show the equally weighted averages of ratios over
velocity ranges in magenta. 
All the resulting OPR averages are listed in Table~\ref{Table: opr results}.

In summary, we find a  value  
above the statistical limit   in the molecular envelope  of W49N, 3.5$\pm0.1$ (formal errors), while for the other three molecular envelopes we find values
slightly below three, 
in the range  \mbox{$(2.3-2.7)\pm0.1$}.  
    In the translucent interstellar gas towards W31C we   find similar values of  \mbox{$(2.2-2.9)\pm0.2$}. 
  However, we also obtain values above three in the translucent gas; towards W31C we 
 find one component with    $\gtrsim 4.2$, and, similarly, towards W49N  one component with 
 $\gtrsim 5.0$. In addition, we find an OPR  of 3.4$\pm0.1$ 
in the redshifted dense and cold filament interacting with W51 at $V_\mathrm{LSR}\sim68$~km~s$^{-1}$.

\subsection{Uncertainties} \label{section: uncertainties}
\begin{figure}[\!htb]  
   \resizebox{\hsize}{!}{
\includegraphics{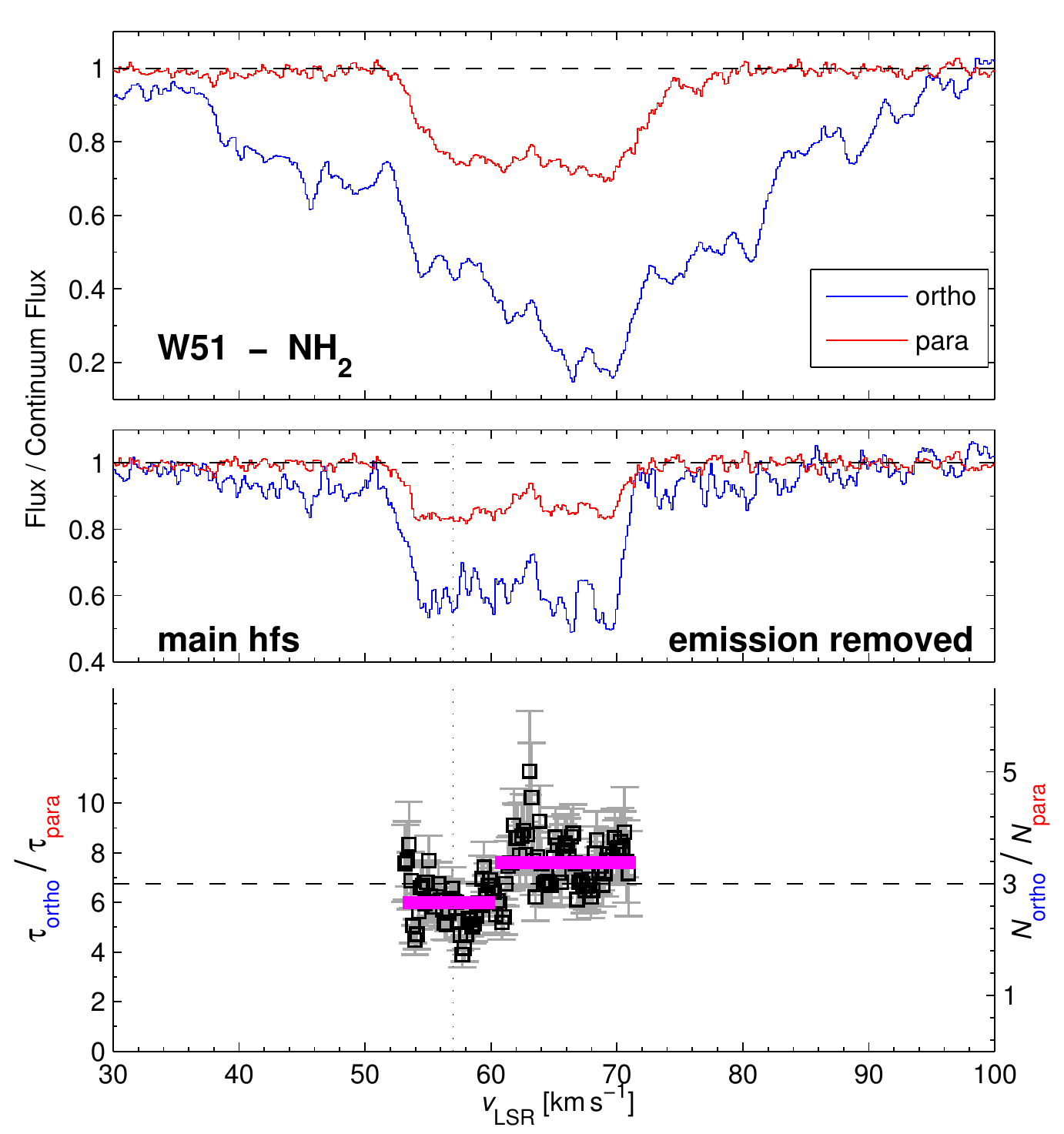}}  
\caption{W51: Notation as in Fig.~\ref{Fig: W31C Orto and para NH2}. 
} 
 \label{Fig: W51 Orto and para NH2}
\end{figure}

\begin{figure}[\!htb]  
 \resizebox{\hsize}{!}{
\includegraphics{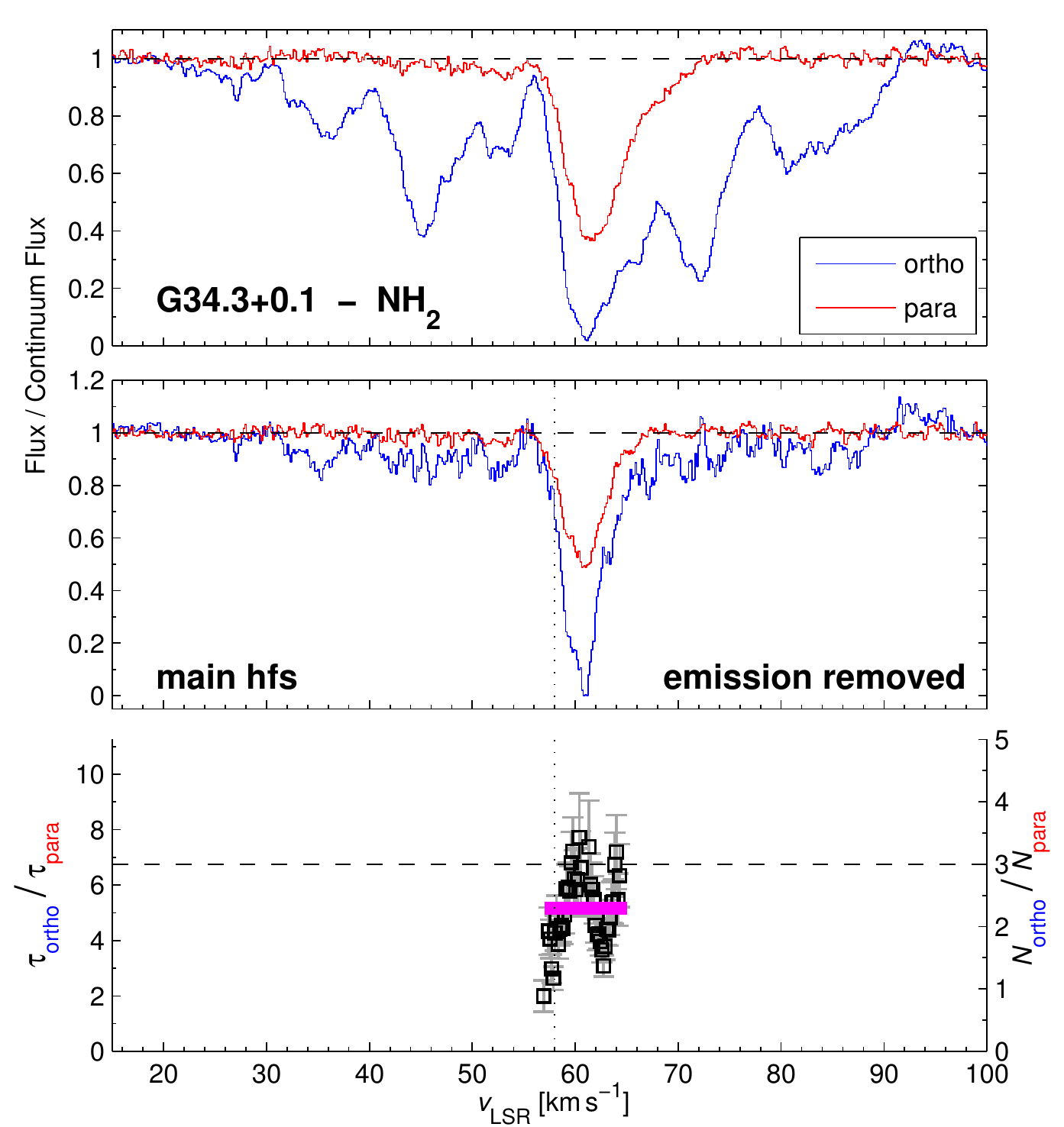}}  
\caption{G34.3+0.1: Notation as in Fig.~\ref{Fig: W31C Orto and para NH2}. 
}
 \label{Fig: G34 Orto and para NH2}
\end{figure}

The 1$\sigma$ errors shown in Figs.~\ref{Fig: W31C Orto and para NH2}--\ref{Fig: G34 Orto and para NH2}  
and given in Table~\ref{Table: opr results}  
correspond to quadratically summed uncertainties that are due to thermal noise and and calibration, including uncertainties 
in the gain ratios of upper and lower sidebands. The variation within the dynamical velocity components seems to be somewhat larger
than the formal errors when averaging over components. We do not know at
this time if this is a real chemical effect or a symptom of imperfections
in our approach or data.

Additional uncertainties, much more difficult to  estimate accurately, 
consist of errors in the deconvolution, emission from the background hot cores, and the
excitation. 
The deconvolution was checked by comparison of the  main hfs component of the respective line obtained from the deconvolution 
with the results from Gaussian fits with good agreements. Examples are shown
for W31C in \mbox{Figs.~\ref{W31C ortho comparison deconv-gaussian fits}-\ref{W31C para comparison deconv-gaussian fits},}  
where we have used the results from the Gaussian fitting performed in 
paper~II.  
The emission of the ortho line from the background hot core has a minor impact on the derived
OPR. For example, in W31C we obtain an OPR of 1.6 without any of the emission 
or correction for excitation. Adding the removal of emission the    mean OPR increases to 1.8.

\begin{table*}[\!htb] 
\centering
\caption{Resulting average OPRs,  observed temperatures, $T_\mathrm{K}$, and the  temperature ranges in which the  OPRs   are reproduced by   model~b,   $T_\mathrm{mod}$, for 
a density of n$_{\rm H}=2\times10^{4}$~cm$^{-3}$ (Fig.~\ref{fig:NH2-ROP_ture}) in the molecular envelopes and the dense core associated with W51,  and 
n$_{\rm H}=1\times10^{3}$~cm$^{-3}$ for the translucent gas (Fig.~\ref{fig:NH2-OPR-translucent}). Details about the models are found in Sect.~\ref{section: discussion}.}
\begin{tabular} {cccccc} 
  \hline\hline
  Source  & $V_\mathrm{LSR}$ & OPR & $T_\mathrm{K}$ & $T_\mathrm{mod}$($t\simeq5\times10^5$~yrs) & $T_\mathrm{mod}$($t\gtrsim10^6$~yrs) \\
  &(km~s$^{-1}$)  & & (K)& (K) & (K)\\
  \noalign{\smallskip}
  \hline
  \noalign{\smallskip}  
  Molecular envelopes \\ 
  \noalign{\smallskip}  
  W31C & -3.5 & 2.5($\pm0.1$) &$30-50$\tablefoottext{a}&$28-35$&$32-35$ \\
  W49N & +8 & 3.5($\pm0.1$)&$\sim$130\tablefoottext{b}&$5-12$&$23-25$ \\   
  W51 & +57  & 2.7($\pm0.1$)&20-50\tablefoottext{c}&$23-28$&$29-32$ \\   
  G34 & +58 & 2.3($\pm0.1$)&20-70\tablefoottext{c,d}&$\gtrsim$35&$\gtrsim$35 \\   
  \noalign{\smallskip}
  \hline
  \noalign{\smallskip} 
  Dense$$ and cold core \\
  \noalign{\smallskip} 
  W51 &  +68 &   3.4($\pm0.1$) &10-30\tablefoottext{f}&$10-13$&$23-25$\\
  \noalign{\smallskip}
  \hline
  \noalign{\smallskip} 
  Translucent gas & & & & $T_\mathrm{mod}$($t\simeq10^4$~yrs) & $T_\mathrm{mod}$($t\gtrsim10^6$~yrs) \\ 
  \noalign{\smallskip}  
  W31C & +22 & 2.2($\pm0.2$) & 30-100\tablefoottext{e} &$\gtrsim$27&$\gtrsim$34 \\
  &  +28 & 2.9($\pm0.2$) & 20-100\tablefoottext{e} &$5-16$& $25-29$ \\
  &  +40 & 2.6($\pm0.2$) & 25-75\tablefoottext{e} &$12-27$&$28-34$ \\    
  \noalign{\smallskip} 
&& & & \multicolumn{2}{c}{$T_\mathrm{mod}$($t\gtrsim 5\times 10^4$~yrs)}\\
 W31C &  +10 & $\gtrsim$1.7   & \ldots &\multicolumn{2}{c}{\ldots} \\ 
 &  +17 & $\gtrsim$4.2 & 30-85\tablefoottext{e} &\multicolumn{2}{c}{$17-21$}\\
W49N &  +39 &  $\gtrsim$5.0 & $<$15\tablefoottext{e} &\multicolumn{2}{c}{$14-19$}\\
  &  +63 &  $\gtrsim$1.4  & 20-120\tablefoottext{e} &\multicolumn{2}{c}{\ldots}  \\
 \noalign{\smallskip}   
  \hline 
\label{Table: opr results}
\end{tabular}
\tablefoot{The tabulated errors are   the formal errors (details 
in Sect.~\ref{section: uncertainties}.)
\tablefoottext{a}{\citet{1978ApJ...221L..77F} and \citet{2002ApJS..143..469M}.}
\tablefoottext{b}{\citet{2001A&A...376.1064V}.}
\tablefoottext{c}{\citet{2013A&A...554A..83V}.}
\tablefoottext{d}{Derived from NH$_3$ rotational transitions  \citep{Hajigholi}.}
\tablefoottext{e}{The  excitation temperature of the
C\ion{I} 492~GHz line  \citep{2015A&A...573A..30G}.}
\tablefoottext{f}{Derived from   CN and
NH$_3$ rotational transitions \citep{2014A&A...566A..61M}.}
}
\end{table*}

 The largest uncertainty in the derived OPR is the excitation,  
as shown in Fig.~\ref{Fig: JTex vs Tex}. 
Using Eq.~\ref{opacity}, we find upper limits to the excitation temperatures as follows: 
  \mbox{$T_\mathrm{ex}(\mathrm{ortho}) \lesssim15.1$~K},  \mbox{$\lesssim17.0$~K}, 
    \mbox{$\lesssim17.4$~K},  and \mbox{$\lesssim10.5$~K}, and 
      \mbox{$T_\mathrm{ex}(\mathrm{para}) \lesssim29$~K},  \mbox{$\lesssim34$~K}, 
    \mbox{$\lesssim33$~K},  and \mbox{$\lesssim26$~K},
    in the  W31C, W49N, W51, and G34.3 molecular envelopes, respectively. 
Applying the correction from the ALI modelling in W31C,  
\mbox{$T_\mathrm{ex}(\mathrm{ortho}) = 12.8$~K} and \mbox{$T_\mathrm{ex}(\mathrm{para}) = 17.4$~K},  
we find that the mean OPR increases from 1.8 
 to 2.5. 
It is clear that if  the excitation temperatures are underestimated, the derived OPRs will also be underestimated and 
could reach, or exceed, the thermal equilibrium value.  However, the opposite is also true. The OPRs will be  
overestimated if we apply too high excitation temperatures. 
The result in  the W31C molecular envelope,  \mbox{OPR = 2.5$\pm0.1$}, can, however, be compared to   
 the   ALI model, where we find  a similar value of 2.6 
 (Fig.~\ref{Fig: W31C ali 2 shell model}), in support of an OPR    below three. 
Towards G34.3, we used an ortho excitation temperature that equals the upper limit obtained from the observed 
 almost saturated line, which   supports a mean OPR below three in this source as well, even though the para excitation may also
 play a role.
 For W49N and W51  the excitation temperature is more difficult to pinpoint, since we were not able to 
 find good ALI models, and the upper limits are rather high. 
The OPR in these sources may in fact have been overestimated since we have 
 used the $T_\mathrm{ex}$  derived from the 
 ALI models  that did not reproduce the depth of the ortho absorptions,
 partly because the excitation temperature was too high.

Assuming that the excitation along the  sight-line gas is low, the OPR in these components is not affected
and  hence can be considered as more robust than the results in the molecular envelopes. 
However, \citet{2013ApJ...765...61E} showed that the assumption that all population of water is in the ground state 
is not valid in the foreground gas towards NGC\,6334I.  The \mbox{ortho-NH$_2$} 953~GHz line is less
affected by the excitation than  the \mbox{ortho-H$_2$O} 557~GHz line, however. 
\citet{2013ApJ...762...11F} studied the water OPR along the same sight-lines as analysed in this paper and
found from analysing the  two ortho ground-state transitions at 557~GHz and 1\,669~GHz that $T_\mathrm{ex}\approx5$~K.  
Assuming that NH$_2$ and NH$_3$ co-exist, we checked in addition our  
ammonia  data along the same sight-lines as reported in this paper.  We observed ammonia \mbox{$J_K = 1_0$-0$_0$}, \mbox{2$_0$-1$_0$}, \mbox{3$_0$-2$_0$}, \mbox{2$_1$-1$_1$}, \mbox{3$_1$-2$_1$}, and \mbox{3$_2$-2$_2$}   lines   
and find no signs of absorptions,   except for the lines connecting to the lowest ground states.   
This suggests that the excitation is also negligible  in the 953~GHz  and 1\,444~GHz   NH$_2$ lines 
\citep[see][for ALI modelling of ammonia in G34.3]{Hajigholi}.

 
\section{Chemistry and ortho-to-para relation of NH$_2$} \label{section: discussion}
In this section we apply a gas-phase chemistry model  
that takes the nuclear-spin symmetries of  molecular hydrogen and the nitrogen hydrides into account. 
The model was  developed for cold and dense gas conditions, however,
which are representative of our molecular envelopes,  
and  it therefore cannot fully exploit the chemistry in the translucent line-of-sight gas.  Work is in progress to include 
 photodissociation and photoionisation reactions and the effect of variation in $A_\mathrm{V}$, as 
 well as surface chemistry.
 
  \subsection{OPR in equilibrium}
 The OPR  can be expressed as the ratio of the sum of the
 populations of the energy levels of ortho-NH$_2$ to those of
 para-NH$_2$. If interconversion between ortho and para states is
 possible and efficient, at local thermodynamical equilibrium (LTE),
 the populations of the energy levels follow a Boltzmann
 distribution. The OPR  is then computed by  the
   ratio of partition functions of the ortho and para forms as 
\begin{equation}
{\rm OPR}(\tkin) = \frac{3 \sum^{ortho}_J \,g_J \,\exp(-E_J/kT)}{\sum^{para}_J \,g_J \,\exp(-E_J/kT)} \approx 
    \exp\left(\frac{30.4}{\tkin}\right) ,
\label{eq:fmuleROP_lowT}
\end{equation}
where $E_J$ stands for the energy of the rotational levels (even though rotation is handled by 
the three quantum numbers $J$, $K_a$, $K_c$ in an asymmetric top). The fine-structure and 
hyperfine-structure contributions to the energy are omitted for simplicity.  
The rotational degeneracy is denoted by $g_J$.
The approximation is valid for low temperatures  \mbox{($T \lesssim 20$~K)}, where only 
the ground states are
 populated  and the partition functions are reduced to the degeneracies
 of the lowest rotational state in each form, which are equal (ignoring the fine and hyperfine structure splitting).  
The energy difference between the
two ground spin states is 30.4~K (see Fig.~\ref{NH2 energy level diagram}).   
 As illustrated in Fig.~\ref{fig:NH2-ROP_ture}, the OPR in thermal equilibrium 
increases with decreasing temperature.

\subsection{Deviation from thermodynamical equilibrium}
\label{subsection:model_a}

\begin{figure}
 \centering
\resizebox{\hsize}{!}{
\includegraphics {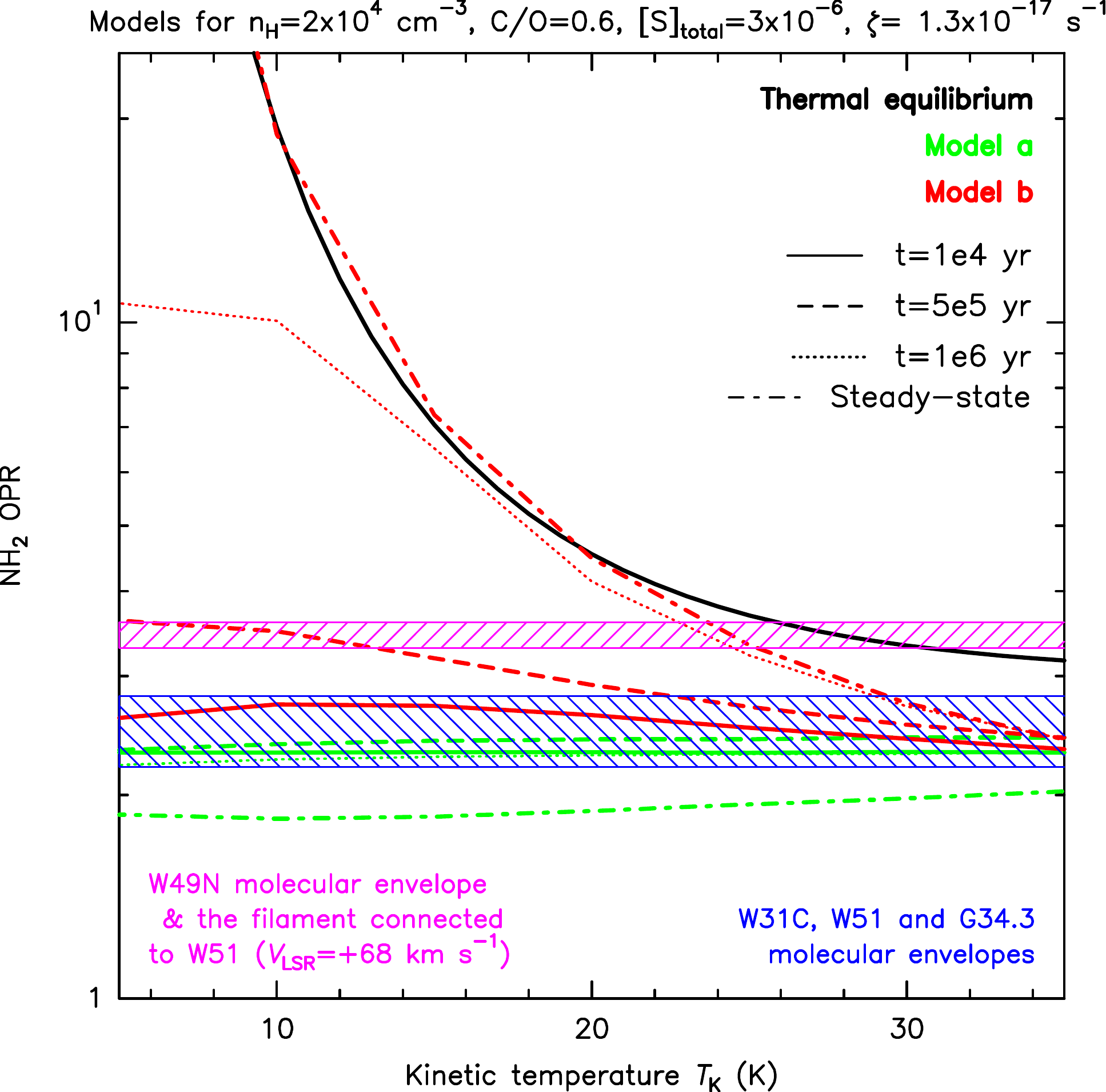}}  
\caption{OPR of NH$_2$  computed as a function of kinetic temperature for a density of
$n_\mathrm{H} = 2\times10^4$~cm$^{-3}$: {\it (i)} at
  thermal equilibrium ({\it black solid line}); {\it (ii)} with  
  {\it \textup{model a}} {\it (green}, Sect.~\ref{subsection:model_a}); and
  {\it (iii)} with {\it \textup{model b}} ({\it red}, corresponding to {\it \textup{model
    a}} plus the \mbox{NH$_2$ + H} reaction,
  Sect.~\ref{subsection:NH2therm}). The OPR is plotted at
 four different times for both models: $1\times10^4$~yr
  (solid lines), $5\times10^5$~yr (dashed lines),  $10^6$~yr
  (dotted lines), and  steady-state (dashed-dotted lines). The
  hatched boxes represent our best estimates of the average OPR,
  including formal error bars. In \textup{{\it \textup{blue}} }we show the W31C, W51,
  and G34.3 molecular envelopes, and in \textup{{\it \textup{pink}}} the filament
  connected to W51 at \mbox{$V_{\rm LSR}\sim68$~km~s$^{-1}$} and the
  W49N molecular envelope.}
 \label{fig:NH2-ROP_ture}
\end{figure}

The pink hatched box in Fig.~\ref{fig:NH2-ROP_ture} shows the range
of observed OPR values, including  their formal 
errors,  in the molecular envelope  of  W49N and 
the cold and dense filament connected to W51 at $V_\mathrm{LSR} \sim 68$~km~s$^{-1}$.  Both OPR values lie above three, while the 
blue hatched box
shows the range
of observed OPR values in the molecular envelopes of  W31C,
W51, and G34.3.  These values lie below three.   Measurements of an OPR lower than its
thermal equilibrium  value in the molecular envelopes
can be due to the non-LTE OPR of H$_2$ pertaining at the low
temperatures of these environments.  
Studies of the OPR of H$_2$ in dense and cold  gas 
\citep{1991A&A...242..235L, 2000A&A...360..656L, 2001ApJ...561..254T, 2006A&A...449..621F, 2009A&A...494..719P} 
have  indeed shown that at low temperatures ($<20$~K) the
OPR  is controlled by kinetic and not by thermodynamical
considerations. At 10~K, model estimates predict an OPR of about
10$^{-3}$ \citep[][]{1991A&A...242..235L, 2006A&A...449..621F}, far
higher than the equilibrium value of $3\times10^{-7}$. Moreover,
recent observations of multi-hydrogenated species such as NH$_3$
\citep{2012A&A...543A.145P} and H$_3^+$ \citep{2011ApJ...729...15C}
show values that significantly depart from thermal equilibrium.
Several measurements of the water OPR have been made, and most indicate values
close to the LTE value of three  \citep{2013ApJ...765...61E,  2013ApJ...762...11F},  although exceptions have been
observed.    \citet{2013JPCA..117.9661L} found an average OPR  along the sight-line towards Sgr\,B2\,(N) of 2.34$\pm$0.25, 
indicating a spin temperature of 24-32~K in thermal equilibrium. 
\citet{2013ApJ...762...11F} found values of 2.3$\pm$0.1 and 2.4$\pm$0.2
 in the +40 and +60~km~s$^{-1}$ components towards W49N,   respectively, 
indicating a spin temperature of $\sim$25~K.   
Other species show OPRs consistent with LTE values,  for example 
H$_2$O$^+$ \citep{2013JPCA..117.9766S, 2013JPCA..11710018G} 
and 
H$_2$Cl$^+$  
\citep{2013JPCA..11710018G, 2015ApJ...807...54N} within the error
bars. 
 
The OPR of H$_2$ has a major impact on the key reaction initiating the
nitrogen hydride synthesis in cold gas conditions, \mbox{N$^+$ + H$_2
  \rightarrow$ NH$^+$ + H}, which possesses a small endoergicity of
the order of the energy difference between the lowest rotational
fundamental states of ortho- and para-H$_2$
\citep[$\sim$170~K,][]{2008Gerlich}. A succession of hydrogenations of
the NH$^+_n$ ions (with $n=1-3$) follows, leading to the formation of
the ion ammonium (NH$^+_4$). Once formed, NH$^+_4$ undergoes
dissociative recombination with electrons resulting in the production
of NH$_3$ and NH$_2$ \citep[see Fig.~3 in][]{2014A&A...567A.130P}.
NH$_2$ can also form through the dissociative recombination of
NH$_3^+$. This pathway, which is not the dominant one for dense and
cold gas, will be more efficient for diffuse and translucent gas,
where the electron fraction can be relatively higher, and
  hence the OPR of NH$_2$ will directly depend
  on the OPR of NH$_3^+$ in these environments. However, since NH$_4^+$ is itself formed
  through NH$_3^+$ + H$_2$, directly or indirectly, the OPR of NH$_2$
  will always depend on that of  NH$_3^+$. 

Considering that the electron recombination rates are equal for the
different nuclear spin species, it is thus the proportion of each form
of NH$^+_4$ (para ($I=0$), ortho ($I=1$), and meta ($I=2$)), 
and also to a lesser extent each form of NH$^+_3$ (depending on the properties of the gas
  considered), that will impact the OPR of NH$_2$. 
 In the cold interstellar medium, the gas-phase production of
NH$_2$ therefore mainly proceeds from NH$^+_4$, while  it is
destroyed by atomic carbon and
oxygen.  Since the destruction rates for the ortho and para forms are
considered to be equal, the OPR of NH$_2$ is  mainly 
determined by the nuclear spin branching ratios in the electron
recombinations of the three spin configuration of NH$^+_4$
\citep{2013JPCA..117.9800R, 2014A&A...562A..83L} {\it viz.,}
  \begin{equation}
\rm OPR =\frac{2 \,m/p(NH^+_4) \, + \,
  \frac{4}{3}\rop(NH^+_4) \, + \, 1}{\frac{2}{3}\rop(NH^+_4) \, + \, 1}\ .
\label{eq:opNH2-dependence}
\end{equation}
Photodissociation of NH$_3$ in diffuse and translucent gas can also form 
NH$_2$, but this pathway is not expected to have a major impact on our results.
The selection rules should, moreover, be the same as those for the reaction NH$_3^+$ + e$^-$. 
Thus, even if the photodissociation of NH$_3$  pathway is the dominant one  in the formation of NH$_2$  
in  diffuse and translucent gas, the effect of the photodissociation of  NH$_3$ will be to 
enhance the NH$_3^+$ + e$^-$ reaction formation pathway outcome in the production of the OPR of  NH$_2$.

To reproduce the observed OPRs presented in this paper, we used
the astrochemical model developed in \citet{2014A&A...562A..83L},
which is based
on gas-phase nuclear-spin conservation chemistry of the
nitrogen hydrides, considering nuclear-spin conservation and
  full scrambling  selection rules for the multi-hydrogenated nitrogen
molecules.  This model\footnote{The model derived from
    \citet{2014A&A...562A..83L} used in this paper has been corrected
    for a typo in branching ratios for the NH$_3^+$ + H$_2$
    reaction. This implies that the observed  OPRs in the range
    2.2 to 2.8, discussed in this paper, are no longer reproduced at
    steady-state, but at earlier times (see
    Fig.~\ref{fig:NH2-ROP_ture}).} 
    (hereafter  model~a)
  can reproduce the observed   OPRs below three 
  in the molecular envelopes of the sources W31C, W51, and G34.3,
  mostly independently of temperature for
  $5<T_\mathrm{K}<35~\mathrm{K}$, for times between $\sim10^{4}$~yr
  and $\sim10^{6}$~yr,  at a density of n$_{\rm H} = 2\times10^{4}$~cm$^{-3}$ 
 and a  cosmic-ray ionisation rate of
    $1.3\times10^{-17}$~s$^{-1}$.   
  These results are represented in green
  in Fig.~\ref{fig:NH2-ROP_ture}. In the on-line appendix Fig.~\ref{fig:NH2-OPR-translucent} 
  we show a model with a lower density, n$_{\rm H} = 1\times10^{3}$~cm$^{-3}$,   and a cosmic-ray ionisation rate of
    $2\times10^{-16}$~s$^{-1}$,   
   which is   compared to our results in the translucent gas.    
   
Model a  cannot reproduce the OPR above three found in the filament
  associated with W51 at $V_\mathrm{LSR}\sim68$~km~s$^{-1}$ ,
however, or the
  limits $\gtrsim4.2$ and $\gtrsim5.0$ along two of the line-of-sight
  components and in the W49N molecular envelope.

\subsection{Possible interconversion between o-NH$_2$ and p-NH$_2$}
\label{subsection:NH2therm}

The ortho and para forms of NH$_2$ can undergo a number of
  exchange collisions with other hydrogenated species such as atomic
  hydrogen, allowing interconversion   between ortho- and para-NH$_2$. 
  Such processes will increase the OPR with decreasing temperature and lead to LTE for the  OPR at steady-state. 
To explain the   OPR values above three, the two
following reactions were added to the \citet{2014A&A...562A..83L} model:
\begin{equation}
{\rm{H}} + {\rm{o-NH_2}}  \stackrel{k_{\rm{o \rightarrow p}}}{\longrightarrow}  {\rm{H}}  + {\rm{p-NH_2}}\ , \label{eq:conversion_onh2_pnh2}
\end{equation}
\begin{equation}
{\rm{H}}  + {\rm{p-NH_2}}  \stackrel{k_{\rm{p \rightarrow o}}}{\longrightarrow}  {\rm{H}}  + {\rm{o-NH_2}}\ , \label{eq:conversion_pnh2_onh2}
\end{equation}
with $k_{\rm{o \rightarrow p}}=k_{\rm{p \rightarrow o}}\exp(-30.4/T)$.
Since these radical-radical  rate coefficients have not yet
been measured or calculated, we chose for $k_{\rm{p \rightarrow o}}$
a typical rate coefficient of $10^{-10}$~cm$^{-3}$s$^{-1}$.  Whether or
  not the H-transfer reactions between H and NH$_2$ proceed at a near
  collisional value is currently uncertain. Experimental evidence from
  the saturated three-body reaction to produce ammonia indicates at
  most a small barrier \citep{pagsberg_pulse_1979}, while theoretical
  calculations indicate a more substantial barrier and a conical
  intersection \citep{mccarthy_dissociation_1987,ma_2012}. New
  calculations  and experiments are clearly needed. 

Results from adding reactions (\ref{eq:conversion_onh2_pnh2}) and
  (\ref{eq:conversion_pnh2_onh2})  to our spin chemistry model, without 
changing any other parameters (hereafter model~b), are represented by the red curves 
in Fig.~\ref{fig:NH2-ROP_ture} and
  \ref{fig:NH2-OPR-translucent}.   
  Since model~b produces 
OPRs below the equilibrium value of three for higher temperatures but approaches the 
LTE ratio with time for low temperatures, it can reproduce all observed OPRs,  
both below and above three. 
At early times, $\sim 10^4$~yr,    the NH$_2$-OPR in  model~b does not have time to thermalise 
and will therefore be close to model~a.  
 As shown in Fig.~\ref{fig:NH2-ROP_ture},  the OPR towards the W49N molecular envelope
  and the filament connected to W51 ($V_\mathrm{LSR} = +68$~km~s$^{-1}$) is  reproduced within a temperature range
  of \mbox{$5-13$~K} at a time equal to $\sim5\times10^{5}$~yr. 
  At later times, the OPR increases more quickly as the temperature decreases, reducing the range of agreement and driving it to higher temperatures. 
  From $10^{6}$~yr, the range remains constant at
  \mbox{$23-25$~K}.  
  The same model parameters can reproduce the OPR range  of $(2.3-2.7)\pm0.1$ in  the envelopes of
  W31C, W51, and G34.3 within a temperature range of
  \mbox{$23-35$~K} at a time of $\sim 5\times10^{5}$~yr as long as the excited
  rotational levels of ortho and para NH$_2$ have a negligible
  population. 
  Finally, the lower limits derived in the translucent gas along the
  sight-lines towards W31C \mbox{($\gtrsim4.2$)} and W49N \mbox{($\gtrsim5$)} can be
  reproduced for temperatures of \mbox{$17-21$~K}, and \mbox{$14-19$~K}, respectively,
  depending on time, using \mbox{n$_{\rm H}=1\times10^{3}$~cm$^{-3}$}   as
  shown in Fig.~\ref{fig:NH2-OPR-translucent}.  
  
  The resulting
  temperatures at different times that can reproduce the observed OPRs
  are consistent with the observed temperatures in the different
  components (modelled and observed temperatures are listed in
  Table~\ref{Table: opr results}), except for the +17~km~s$^{-1}$
  component towards W31C and the molecular envelope in W49N. The
  result in the latter component may suggest that we may have
  overestimated the OPR in W49N by using a too high excitation
  temperature. By using $T_\mathrm{ex}$(ortho) = 12~K instead of
  14~K, for instance, the mean OPR decreases to 3.0. 
  
  We note that the resulting timescales are only indicative
    since our time-dependent models depend on many uncertain physical
    and chemical parameters. For  model b, which reproduces the
     the observed OPRs best, decreasing the density will lower the
    temperature ranges that reproduce the OPRs for specific timescales
    at which the thermal equilibrium is not yet reached. The initial
    chemical conditions can also play an important role in the
    chemistry. For a given density, increasing the ionisation
    rate will moreover have the effect of shortening the timescale so that 
    steady-state will be reached faster.  In addition, for a given time, the
    model will require  higher temperatures to reproduce the observed OPRs.

If the exchange collisions between H and NH$_2$ are sufficiently
reduced in rate, the OPR should lie in between the initial OPR value,
when it forms through exothermic dissociative electronic
recombinations of NH$^+_4$, and the thermalised value depending upon
time \citep[cf.][]{2015ApJ...807...54N}. For the model including
\mbox{H-exchange} reactions between H and NH$_2$ model~b in
Figs.~\ref{fig:NH2-ROP_ture} and \ref{fig:NH2-OPR-translucent}), the
$k_{\rm{o \rightarrow p}}$ rate coefficient used assumes that only the
ground para and ortho states are populated ($0_{00}$ and $1_{01}$
states). However, the first excited ortho ($1_{11}$) and para
($1_{10}$) levels lie only 45~K and 22~K above their respective ground
levels, which implies that the two-level conversion rate is not
accurate above $\sim$20~K if the excited rotational states of ortho
and para are significantly populated.  As suggested in
\cite{2015EPJWC..8406002H}, two points have to be examined. First, if
the density is below the critical density, spontaneous emission
dominates collisional effects. In that case, as can occur in diffuse
gas where the rotational temperature can be quite low, the OPR is
approximated by Eq.~\ref{eq:fmuleROP_lowT}. Secondly, if chemical
equilibrium is not reached, the OPR should be kinetically controlled
(as in Eq.~\ref{eq:opNH2-dependence}). Once formed, the para and ortho
NH$_2$ relax towards lower rotational states for each nuclear spin
state preserving the original OPR. If the relaxation occurs mainly by
spontaneous emission, the lowest rotational states will be formed
predominantly. Afterwards, a competition ensues between spin
equilibration, which in   model b  mostly occurs via exchange
collisions with H atoms, and destruction processes with a variety of
radicals. If destruction dominates, the eventual OPR will be
kinetically controlled. 

As a comparison with the OPR observations of  NH$_3$ already published
  in \cite{2012A&A...543A.145P} and also to highlight the coherence of
  our ortho-para model, we present the results of models~a and 
  b for the OPR values of NH$_3$ in the on-line Fig.~\ref{fig:NH3-OPR-translucent},  where we have used
  the same physical parameters as for the lower density NH$_2$-OPR
  model. The hatched box marks the observed NH$_3$-OPR values along the
  sight-lines towards W31C and W49N \citep{2012A&A...543A.145P}.  
  Our model can reproduce the observed OPRs depending
  on the physical conditions and timescales considered. Increasing the
  cosmic-ray ionisation rate or decreasing the density decreases the
  steady-state value of the NH$_3$ OPR, which shortens the modelled timescales 
 necessary to reproduce the observed OPR values. 

Most of the NH$_2$-OPR values derived from the observations presented in this
paper are consistent with a gas-phase spin nitrogen chemistry
 including H-exchange reactions between H and NH$_2$. We therefore
did not include surface chemistry in our model, except for the
formation of H$_2$ and charge exchanges. Future development of our
model will include the possibility of ortho-para conversion on cold
grain surfaces, which may take place in molecular clouds with low
temperature and high density where gas-phase species condense on the
surface of grains \citep{2002ApJ...569..815T}.  The efficiency of this
phenomenon depends on the time the species reside on the grain, on the
grain surface shape, and on the time of nuclear spin conversion.
However, the characteristic nuclear-spin conversion times on grain
surfaces are not yet well constrained \citep{2000A&A...360..656L,
  2011PCCP...13.2172C,hama_2013}.
Neither did we
consider forbidden spontaneous emission \citep{2013JPCA..117.9584T}
or state-specific formation and destruction in our models, which might also
help to explain the OPR values above three.
In diffuse gas, the UV radiation might also play a significant role in
the nitrogen chemistry by photo-dissociation
\citep{2012A&A...543A.145P}, although we do not expect that UV
radiation significantly changes the OPR of nitrogen hydrides, which is mainly
driven by proton transfer reactions.


\section{Conclusions}
We have derived ortho-to-para ratios of NH$_2$  both above and below the statistical value of three.
In the  molecular envelopes surrounding three of our observed hot cores, we found average OPRs  \emph{below} three 
and similar values in the translucent interstellar gas. 
In contrast, we found an average OPR   \emph{above} three in  the dense and cold filament interacting with W51, in addition to 
two velocity components along the sight-lines towards W31C and W49N and in the molecular
envelope in W49N. The 
results in the line-of-sight interstellar gas are considered to be more robust than in the source molecular clouds 
because of the lower or completely negligible excitation.  
Using a non-LTE radiative transfer model based on the accelerated lambda iteration scheme, we  
successfully modelled 
the emission and absorption in the W31C molecular envelope supporting an OPR below three. 
The derived OPR below three in the G34.3 molecular envelope is also considered as more robust than the
results in the W49N and W51 molecular envelopes, since we used
the upper limit of the ortho excitation temperature when correcting the  opacities for excitation.  

Astrochemical models considering nuclear-spin gas-phase chemistry in a
``para-enriched H$_2$'' gas can reproduce the variations of the
observed ortho-to-para ratios of NH$_2$ and NH$_3$  below their statistical
value of three  and unity, respectively. The NH$_2$ OPR values  
higher than three can be explained by including in these models
the thermalisation reaction \mbox{NH$_2$ + H} with efficient
ortho-to-para and para-to-ortho rate coefficients and selected
intervals of time.

\begin{acknowledgements}
HIFI has been designed and built by a consortium of institutes and university departments from across
Europe, Canada and the United States under the leadership of SRON Netherlands Institute for Space
Research, Groningen, The Netherlands and with major contributions from Germany, France and the US.
Consortium members are: Canada: CSA, U.Waterloo; France: CESR, LAB, LERMA, IRAM; Germany:
KOSMA, MPIfR, MPS; Ireland, NUI Maynooth; Italy: ASI, IFSI-INAF, Osservatorio Astrofisico di Arcetri-
INAF; Netherlands: SRON, TUD; Poland: CAMK, CBK; Spain: Observatorio Astron{\'{o}}mico Nacional (IGN),
Centro de Astrobiolog{\'{i}}a  (CSIC-INTA). Sweden: Chalmers University of Technology - MC2, RSS \& GARD;
Onsala Space Observatory; Swedish National Space Board, Stockholm University - Stockholm Observatory;
Switzerland: ETH Zurich, FHNW; USA: Caltech, JPL, NHSC. 
Support for this work was provided by NASA through an award issued by JPL/Caltech.
CP, ES and MO acknowledge generous support from the Swedish National Space Board. 
R. L. and E. H. acknowledge support for this work provided by NASA
through an award issued by JPL/Caltech.
A.F. acknowledges support from the Agence Nationale de la Recherche (ANR-HYDRIDES), contract ANR-12-BS05-0011-01. 
MG and DL acknowledge support from Centre Nationale d'Etudes Spatiales
(CNES) and the CNRS/INSU programme de physique et chimie du milieu
interstellaire (PCMI).
 
\end{acknowledgements}

\bibliographystyle{aa-package/bibtex/aa}
\bibliography{references}

\Online
\appendix
 
 \clearpage
 \section{Figures}

\begin{figure}  
\centering
\resizebox{\hsize}{!}{
\includegraphics[scale=0.35]{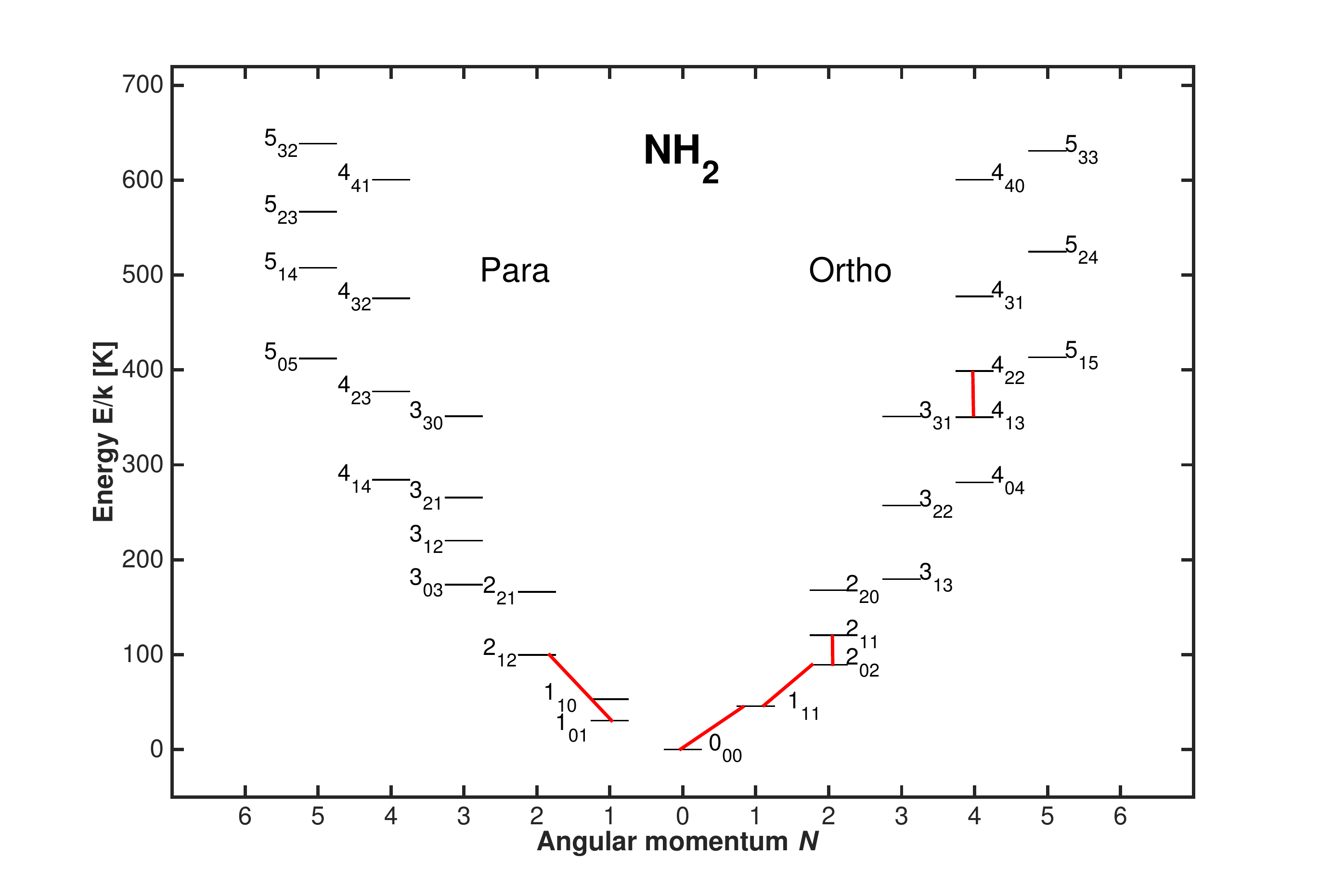}} 
\caption{Energy level diagram of NH$_2$. The observed transitions in this paper are marked in red. 
}
 \label{NH2 energy level diagram}
\end{figure}

\begin{figure}  
\centering
\resizebox{\hsize}{!}{
\includegraphics{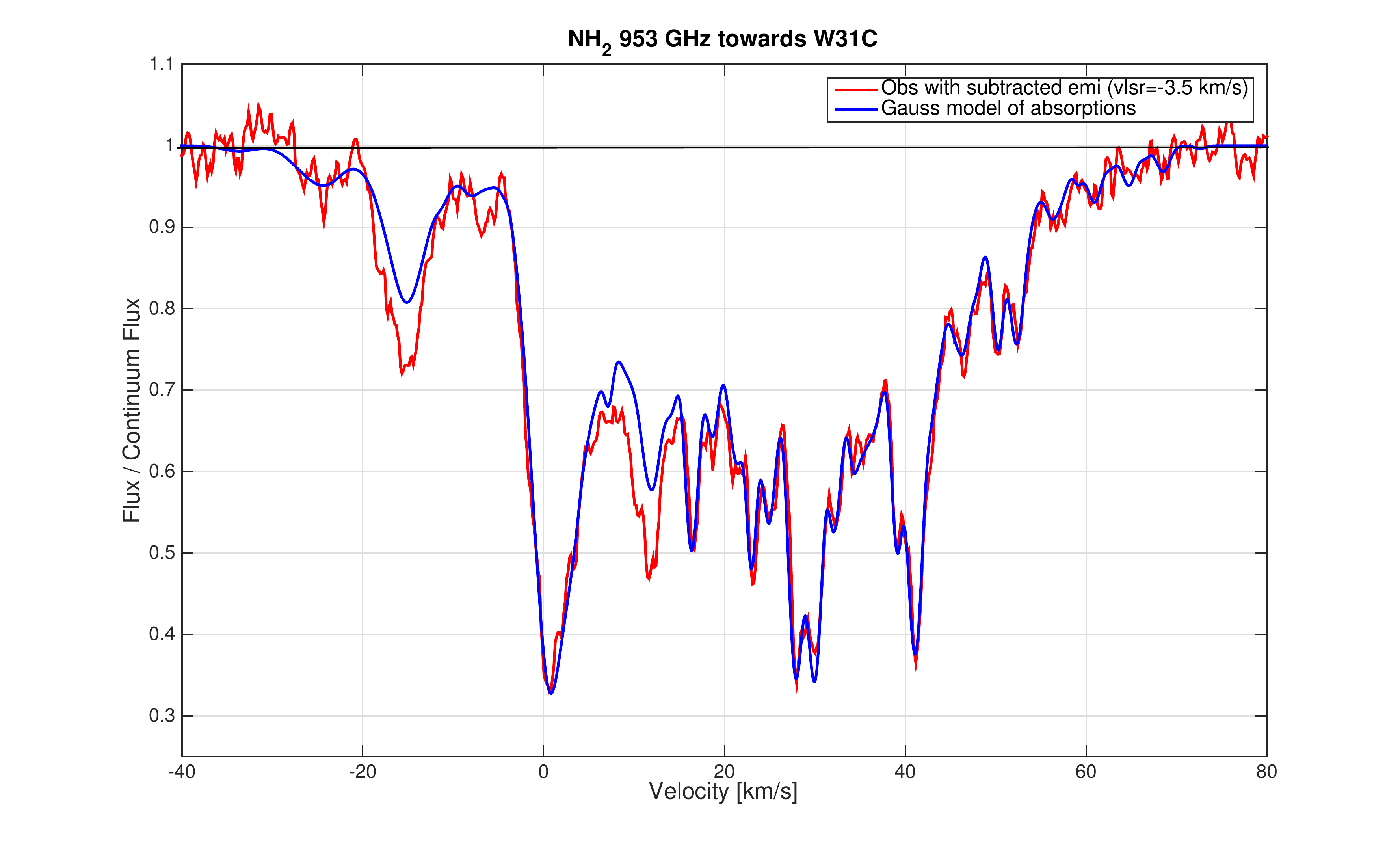}}
\caption{\emph{W31C}. 
\emph{Red}: Original observations (channel separation 0.157~km~s$^{-1}$) with subtracted emission. 
\emph{Blue}:  Gaussian fits of all velocity (and hfs) components. 
Line widths and $v_\mathrm{LSR}$ for each velocity component were taken from the 
results of the  simultaneous Gaussian fitting of NH, ortho-NH$_2$ , and ortho-NH$_3$ (method~I, described in paper~II). 
}
 \label{W31C ortho comparison deconv-gaussian fits}
\end{figure}

 \begin{figure}  
\includegraphics[scale=0.61]{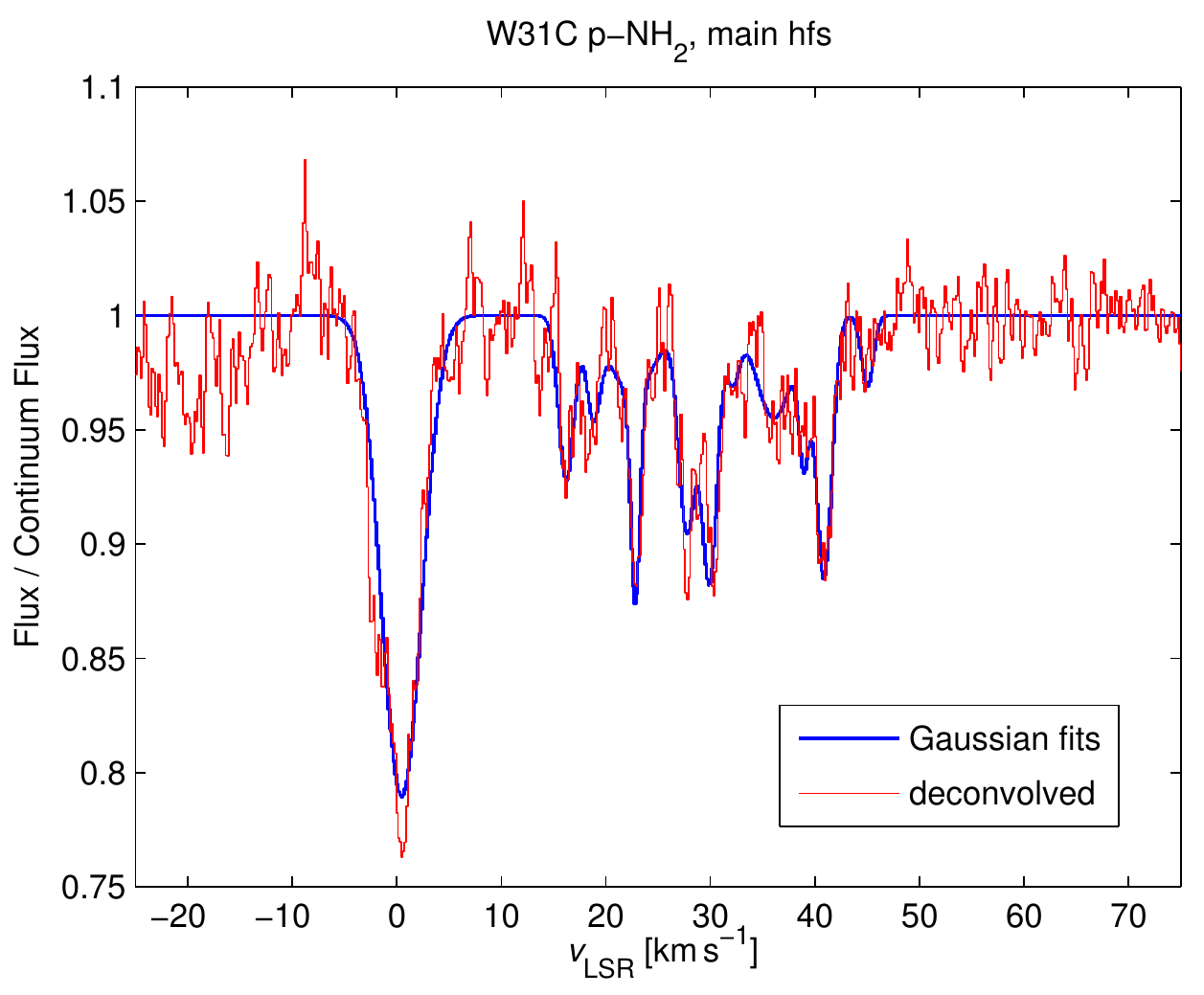}
\caption{\emph{W31C}. 
\emph{Red}: The strongest hfs component of para-NH$_2$ obtained from deconvolution. 
\emph{Blue}: The strongest hfs component of para-NH$_2$ obtained from Gaussian fits of the opacities. 
Line widths and $v_\mathrm{LSR}$ for each velocity component were taken from the 
results of the  simultaneous Gaussian fitting of NH, ortho-NH$_2$ , and ortho-NH$_3$ 
(method~I, described in paper~II). 
}
 \label{W31C para comparison deconv-gaussian fits}
\end{figure}

\begin{figure}  
\resizebox{\hsize}{!}{
\includegraphics{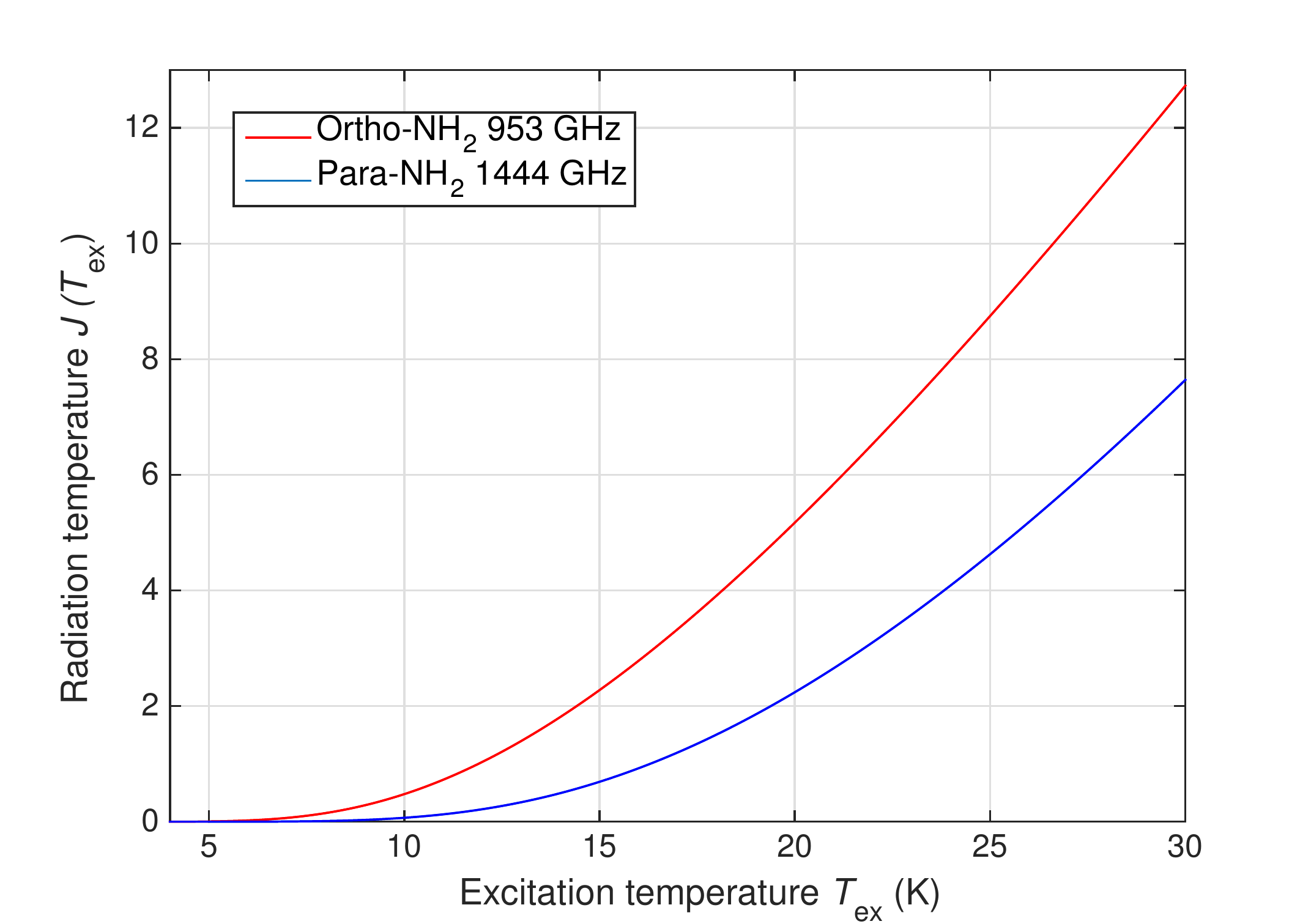}}
\caption{Radiation temperature $J(T_\mathrm{ex})$ as a function of the excitation temperature  $T_\mathrm{ex}$ of the
ortho   $1_{1,1} 3/2 -  0_{0,0} 1/2$ (953~GHz) and para   $2_{1,2} 5/2 -  1_{0,1} 3/2$ (1444~GHz) lines. 
}
 \label{Fig: JTex vs Tex}
\end{figure} 

%
 
 \clearpage
 \begin{figure}  
\resizebox{\hsize}{!}{
\includegraphics{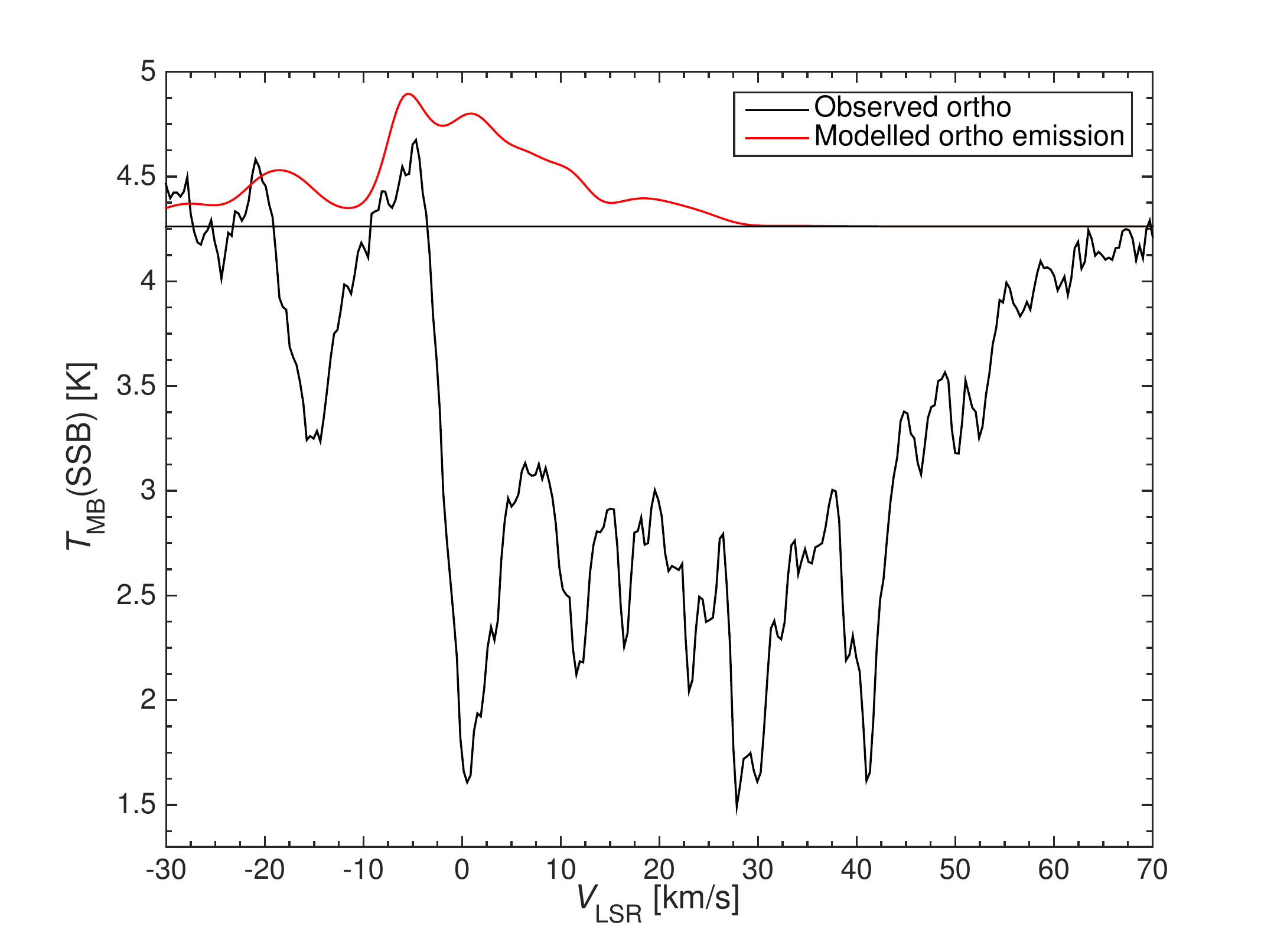}}
\caption{W31C: Modelled ortho emission from the hot core with the ALI code (parameters are listed in
Table~\ref{table: ali parameters}).}
 \label{Fig: W31C ali emission}
\end{figure}

\begin{figure}  
\resizebox{\hsize}{!}{
\includegraphics{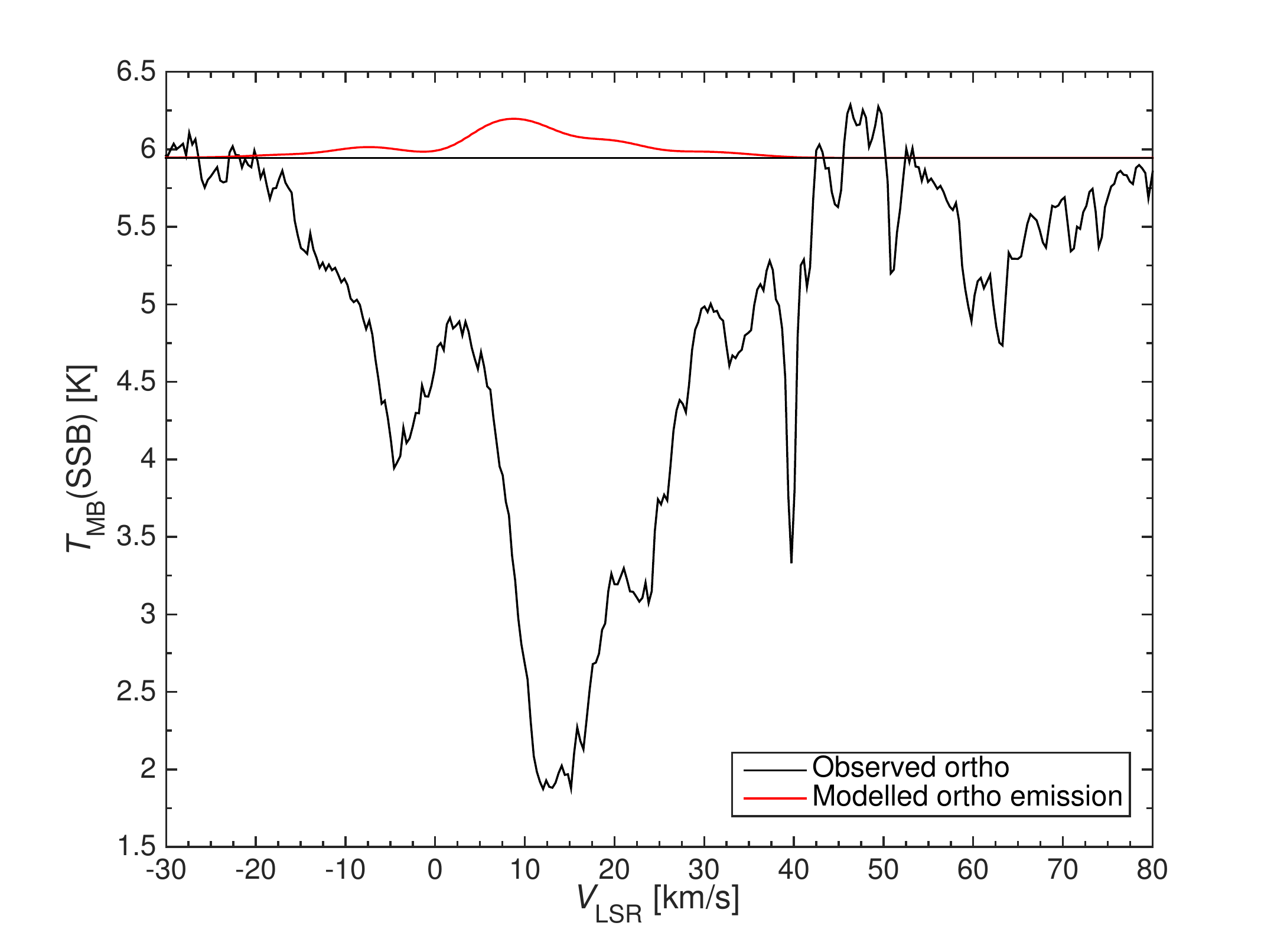}}
\caption{W49N: Notation as in Fig~\ref{Fig: W31C ali emission}.}
 \label{Fig: W49N ali emission}
\end{figure} 

\begin{figure}  
\resizebox{\hsize}{!}{
\includegraphics{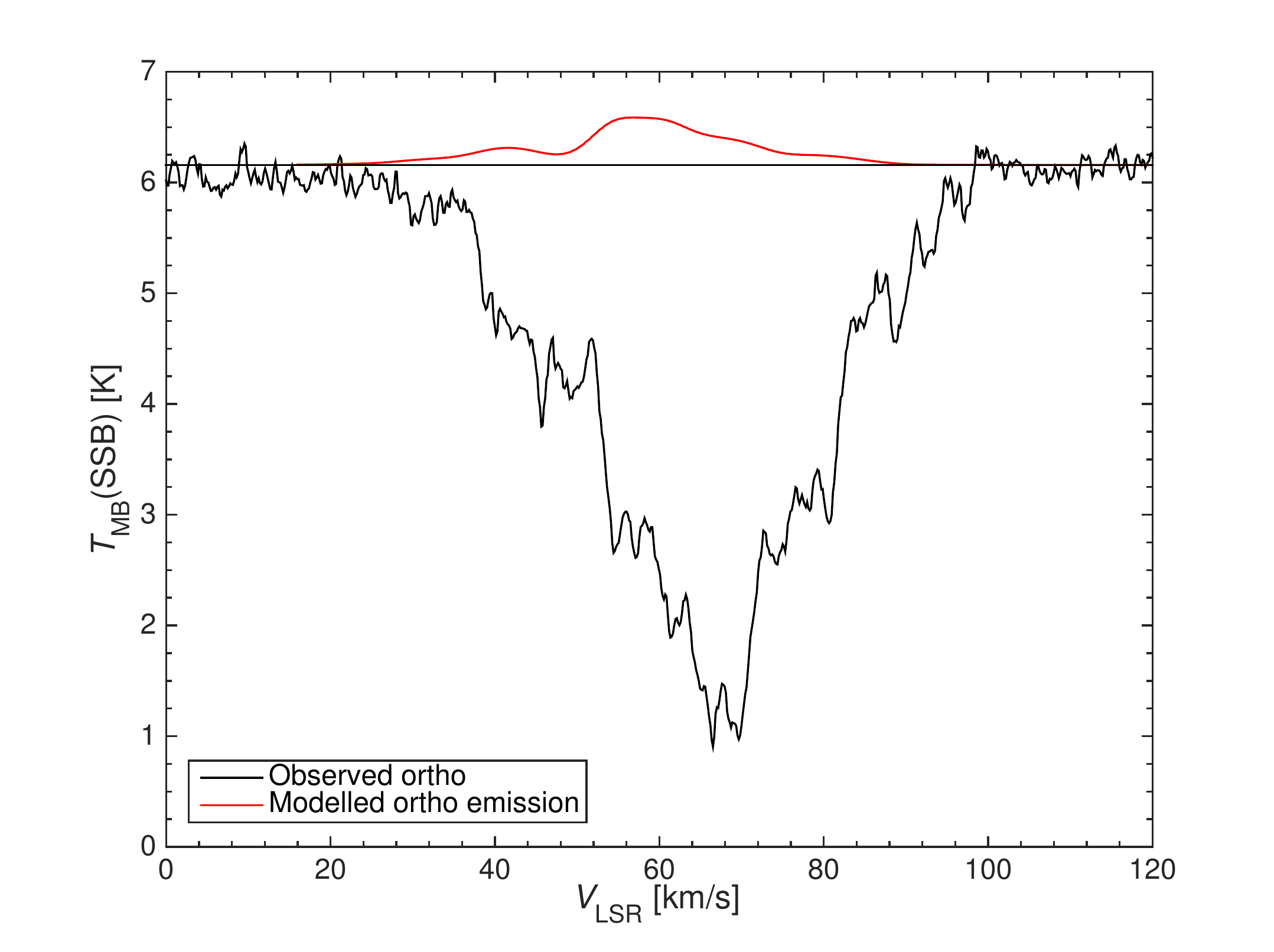}}
\caption{W51: Notation as in Fig~\ref{Fig: W31C ali emission}.}
 \label{Fig: W51 ali emission}
\end{figure} 

 \begin{figure}  
\resizebox{\hsize}{!}{
\includegraphics{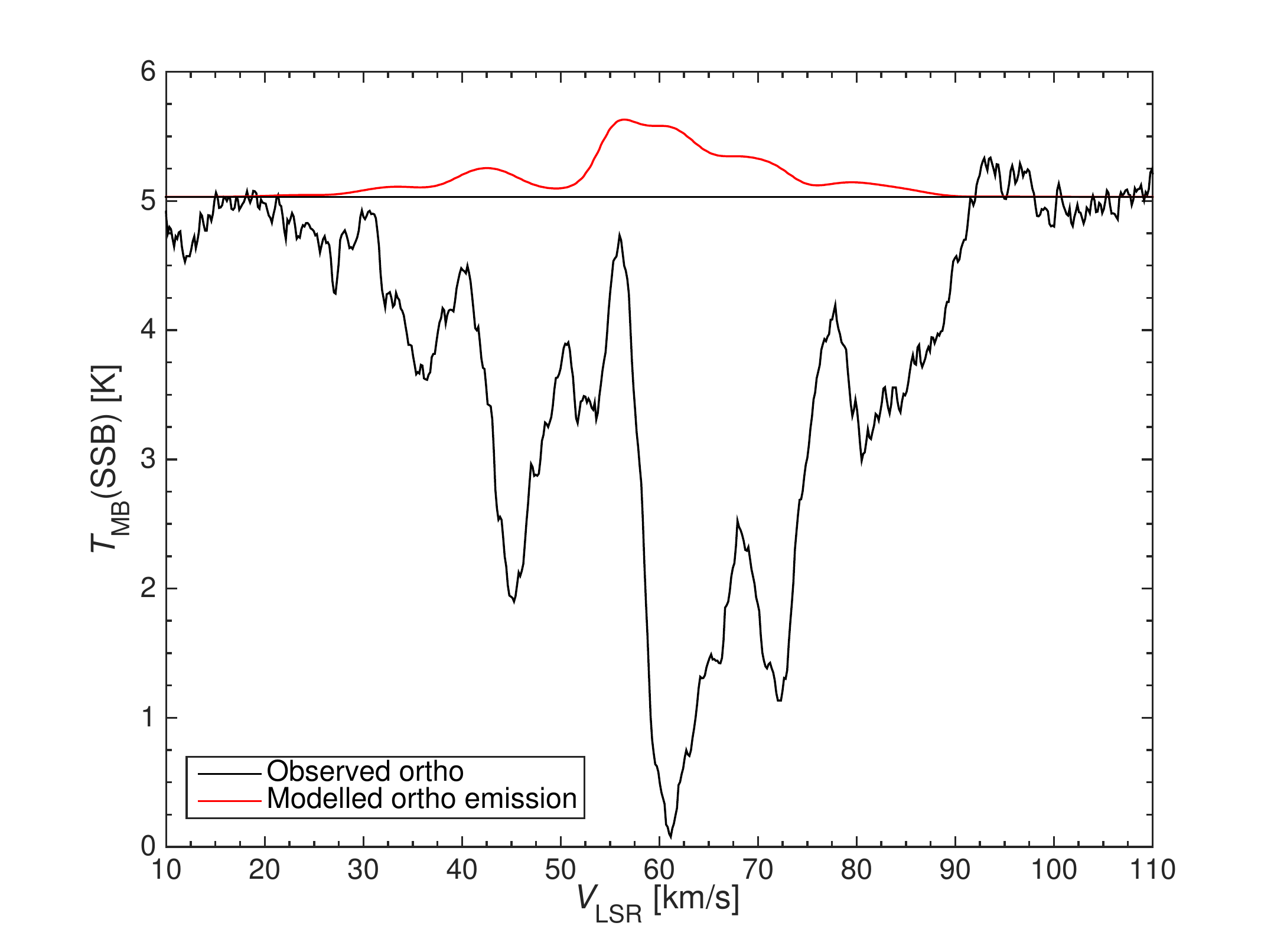}}
\caption{G34.3: Notation as in Fig~\ref{Fig: W31C ali emission}.}
 \label{Fig: G34 ali emission}
\end{figure}

%

\begin{figure} 
  \resizebox{\hsize}{!}{
\includegraphics[scale=0.61]{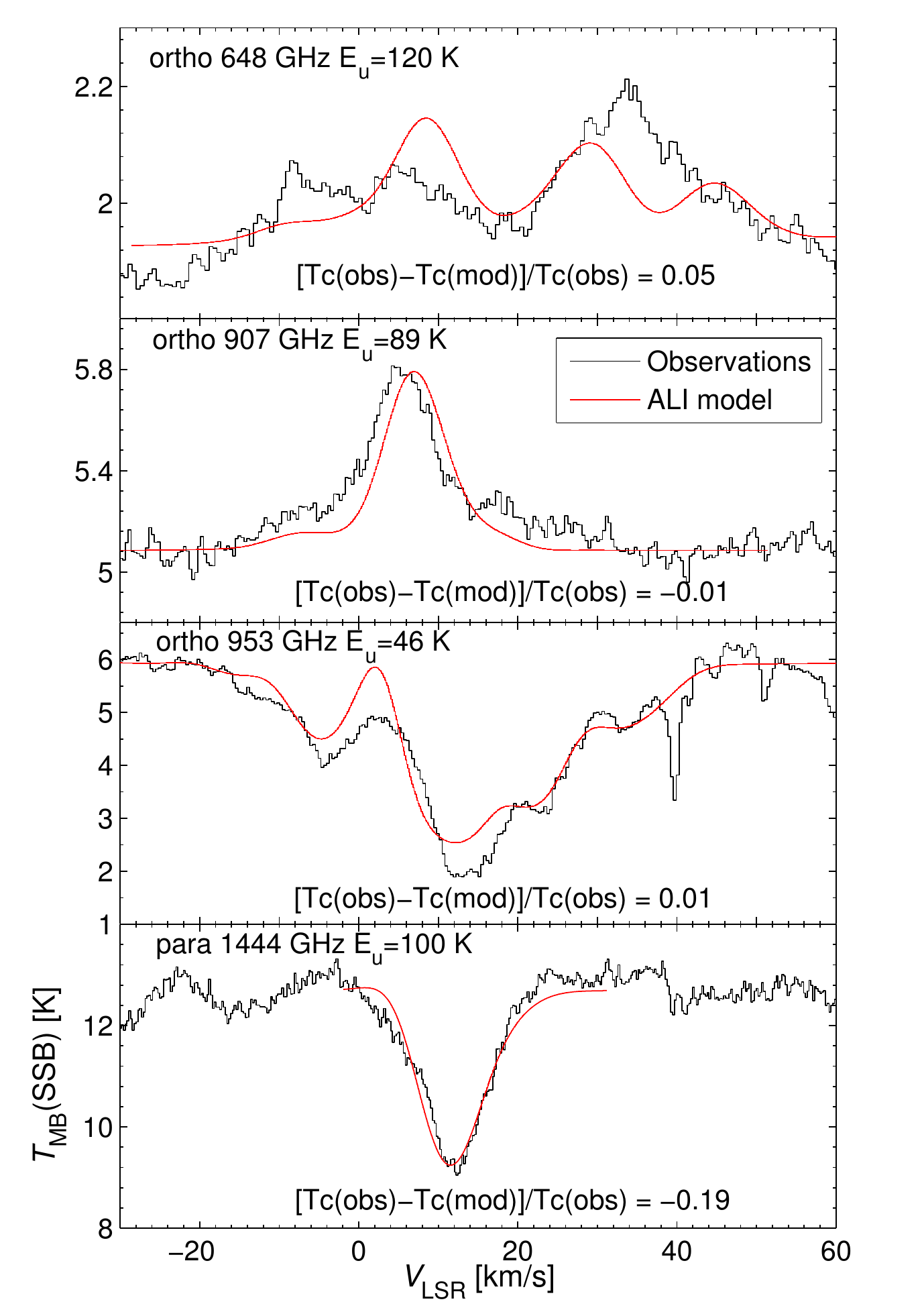}} 
\caption{Example of a simple ALI model of the W49N hot core and the surrounding envelope (parameters are listed in
Table~\ref{table: ali parameters}). 
The relative difference of the modelled  and observed continuum
at the frequency of each line is given in the respective legend.
The 649~GHz line suffers from contamination from the image sideband and is used as an upper limit of
its true intensity. 
} 
 \label{Fig: W49N ali 2 shell model}
\end{figure}

\begin{figure} 
  \resizebox{\hsize}{!}{
\includegraphics[scale=0.61]{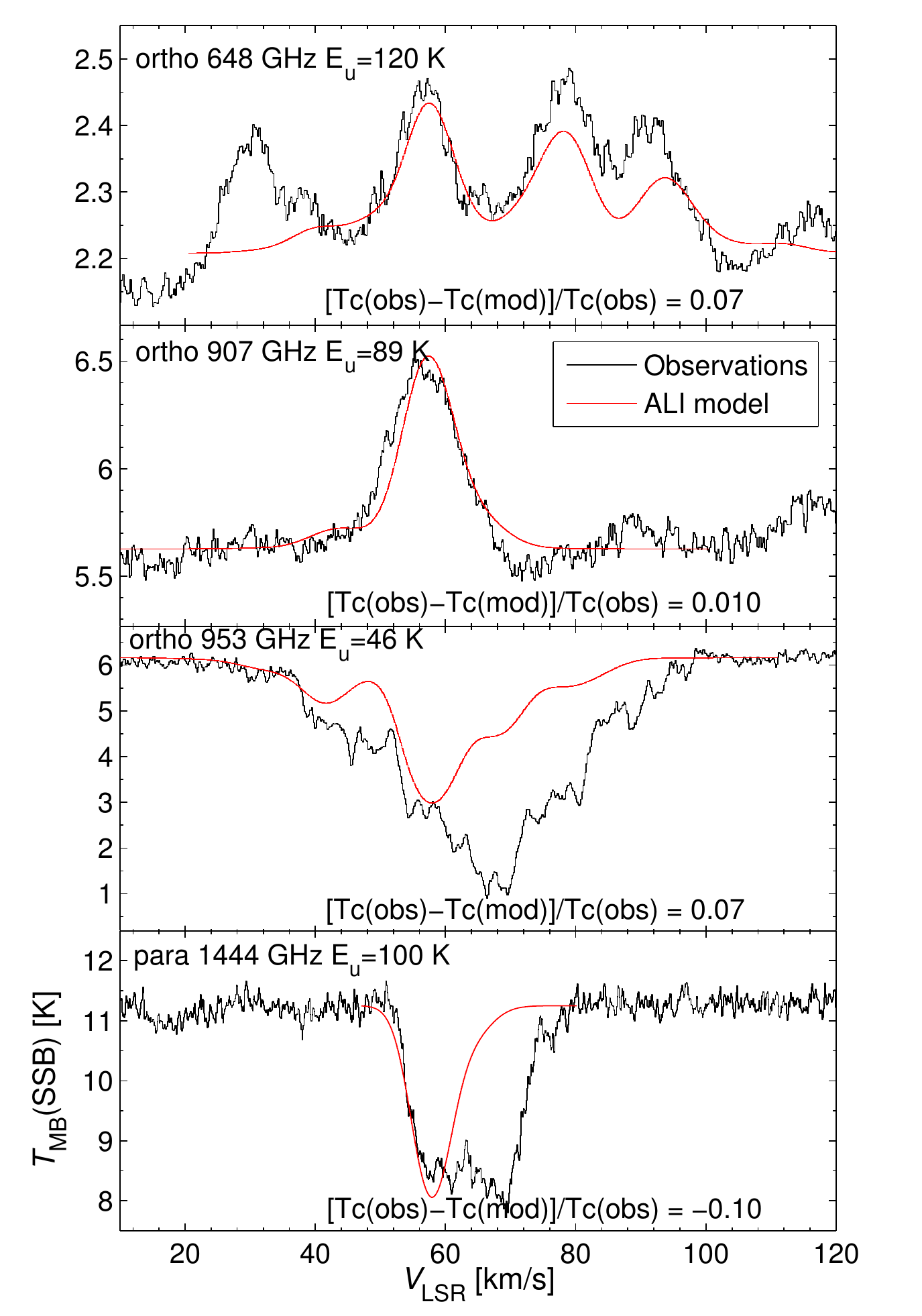}} 
\caption{W51. Notation as in Fig.~\ref{Fig: W49N ali 2 shell model}.  
The foreground absorption centred at $V_\mathrm{LSR} = 68$~km~s$^{-1}$ 
is not modelled.
} 
 \label{Fig: W51 ali 2 shell model}
\end{figure}

 \begin{figure} 
\resizebox{\hsize}{!}{
\includegraphics[scale=0.61]{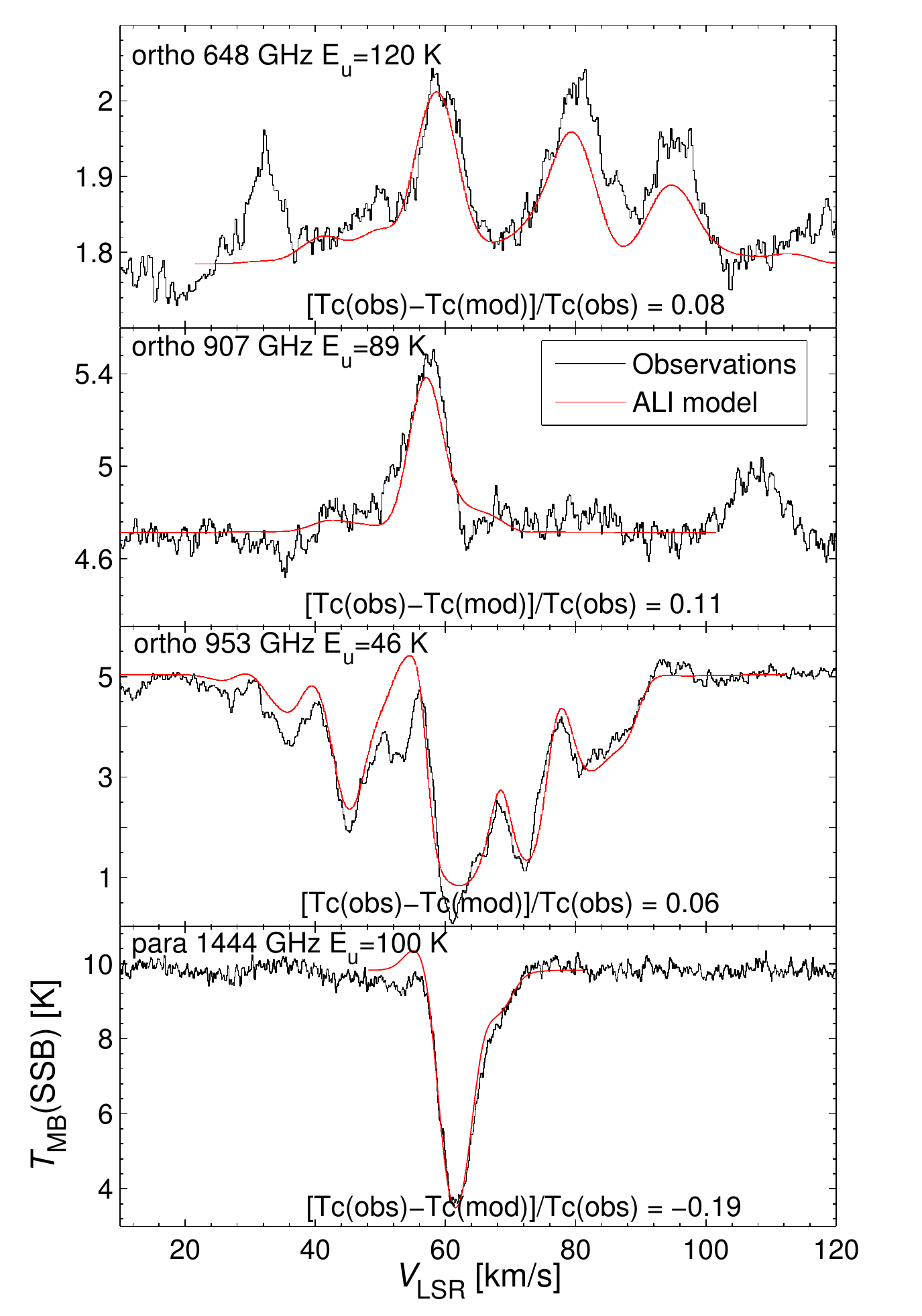}}  
\caption{G34.3+0.1.  Notation as in Fig.~\ref{Fig: W49N ali 2 shell model}.   
} 
 \label{Fig: G34 ali 2 shell model}
\end{figure}

\begin{figure}
 \centering
\resizebox{\hsize}{!}{
\includegraphics{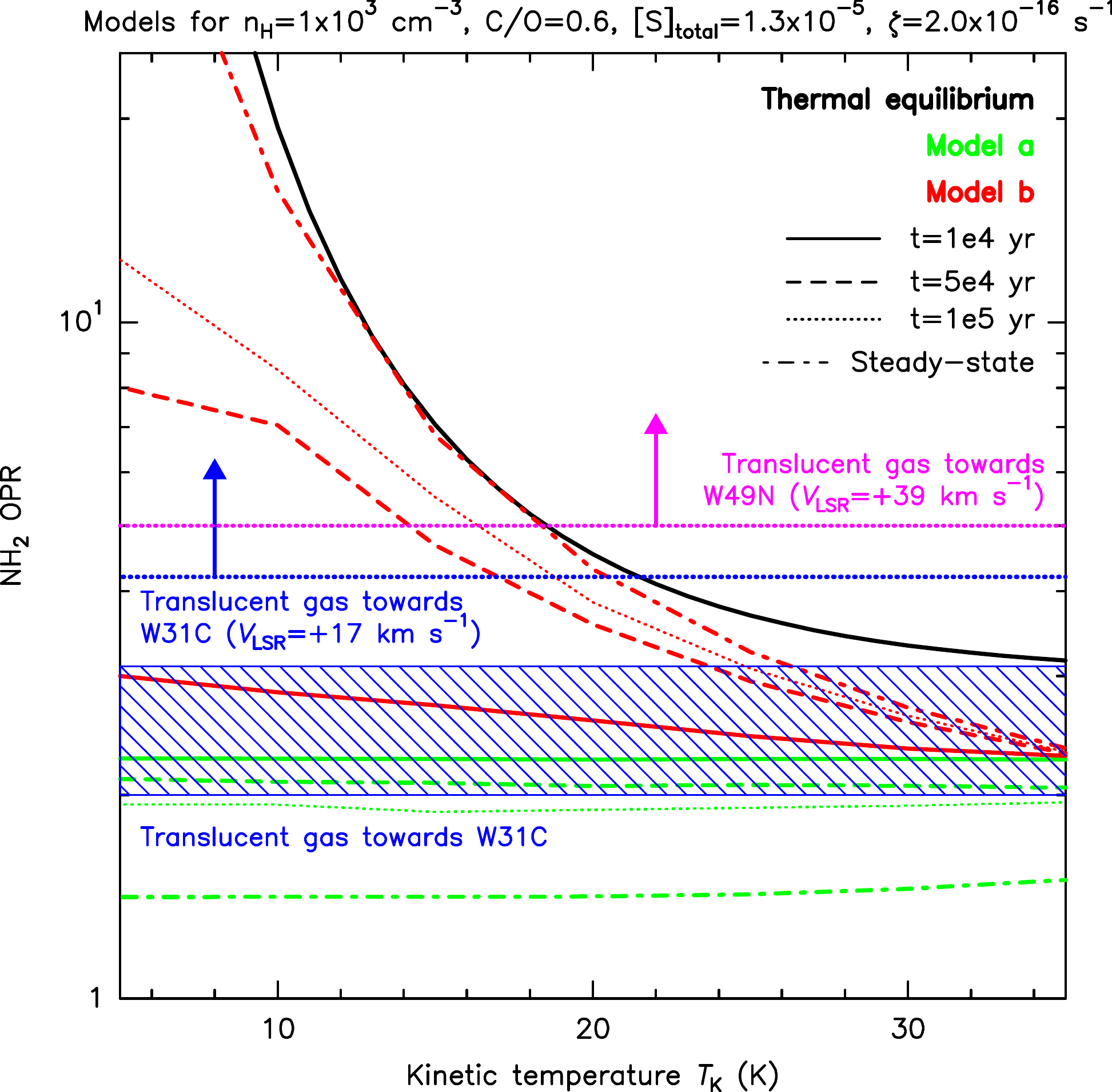}}  
\caption{Calculated OPR  of NH$_2$ shown as a function of kinetic temperature for a density of 
  n$_{\rm H}=1\times10^{3}$~cm$^{-3}$: {\it (i)} at thermal
  equilibrium ({\it black solid line}), {\it (ii)} with 
  {\it model a} {\it (green}, Sect.~\ref{subsection:model_a}); and
  {\it (iii)} with {\it model b} ({\it red}, corresponding to {\it
    model a} plus the \mbox{NH$_2$ + H} reaction,
  Sect.~\ref{subsection:NH2therm}). The OPR is plotted for four
  different times for both models: $1\times10^4$~yr (solid lines),
  $5\times10^4$~yr (dashed lines),  $10^5$~yr (dotted lines), and
  steady-state (dashed-dotted lines). The hatched box
  represents our best estimates of the average OPR, including formal
  errors, for the translucent gas towards W31C at \mbox{$V_{\rm
      LSR}\sim22, 28$~and~$40$~km~s$^{-1}$}. The
  dotted horizontal lines with arrows mark the lower limits in the
  translucent gas towards W31C at \mbox{$V_{\rm
      LSR}\sim17$~km~s$^{-1}$} ({\it blue}) and W49N at \mbox{$V_{\rm
      LSR}\sim39$~km~s$^{-1}$} ({\it pink}).}
 \label{fig:NH2-OPR-translucent}
\end{figure}

\begin{figure}
 \centering
\resizebox{\hsize}{!}{
\includegraphics {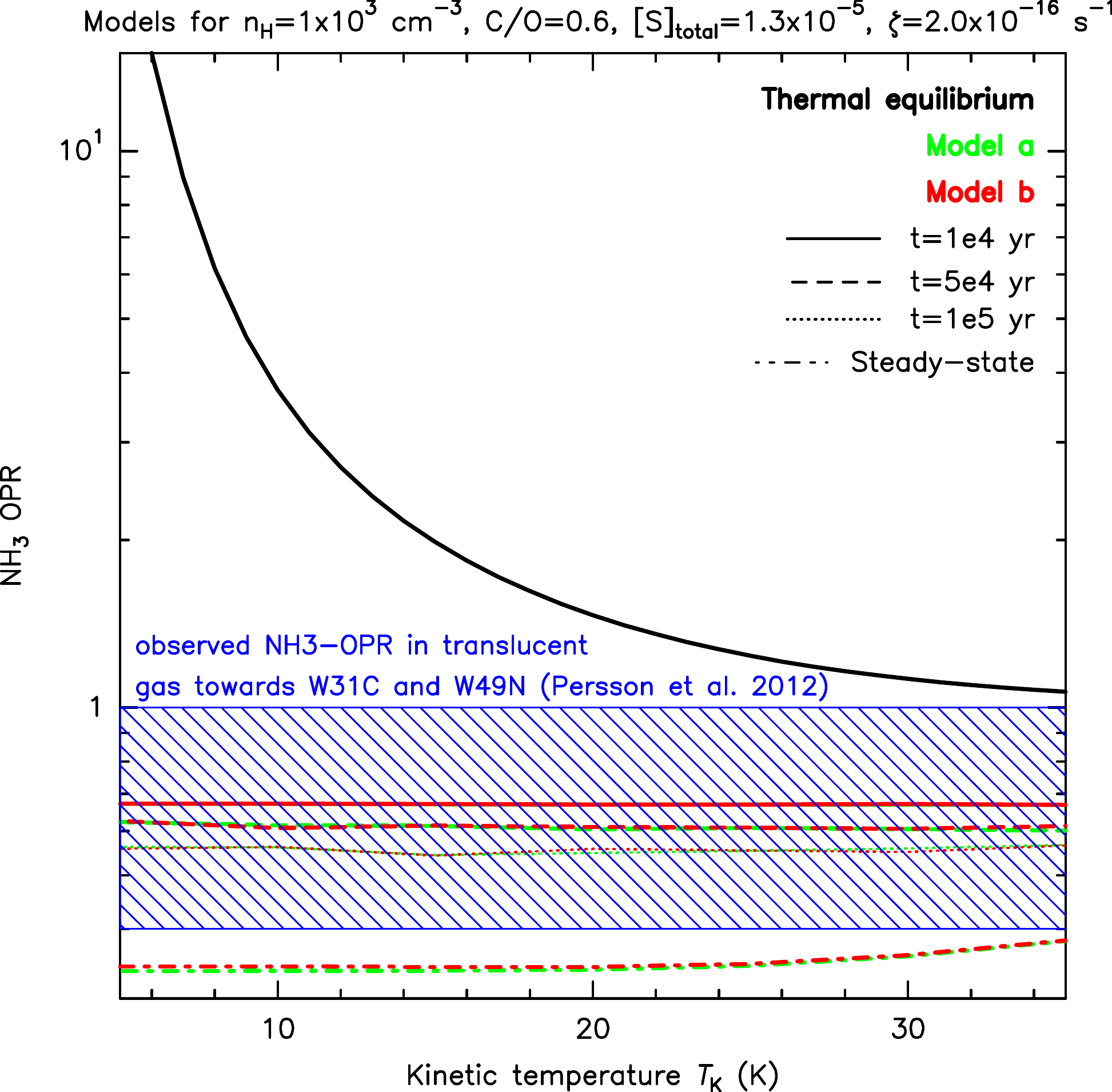}} 
\caption{OPR of NH$_3$ shown as a function of kinetic temperature for a density of 
  n$_{\rm H}=1\times10^{3}$~cm$^{-3}$: {\it (i)} at
  thermal equilibrium ({\it black solid line}); {\it (ii)} with  
  {\it model a} {\it (green)} (details in 
  Sect.~\ref{subsection:model_a}); and {\it (iii)} with
  {\it model b} ({\it red}) (corresponding to  {\it model
    a} plus the \mbox{NH$_2$ + H} reaction, details in 
  Sect.~\ref{subsection:NH2therm}). The OPR is
  shown at four different times for both models: $1\times10^4$~yr (solid lines),
  $5\times10^4$~yr (dashed lines),  $10^5$~yr (dotted lines), and
 steady-state (dashed-dotted lines). The hatched box represents 
  the  OPR of NH$_3$, including formal errors, in the translucent gas in the sight-lines towards W31C and W49N
  from \citep{2012A&A...543A.145P}. 
  }
 \label{fig:NH3-OPR-translucent}
\end{figure}

%
%
 
\clearpage

\section{Tables}

 \begin{table*}[\!htb] 
\centering
\caption{\emph{Herschel} OBSIDs of the observed transitions.
}
\begin{tabular} {llrclccc } 
 \hline\hline
     \noalign{\smallskip}
Source & Species        & Frequency & Band & LO-setting\tablefootmark{a} & Date &        OBSID    \\    \noalign{\smallskip}

&  & (GHz)    \\
     \noalign{\smallskip}
     \hline
\noalign{\smallskip}  

W49N & o-NH$_2$ & 648.784 & 2a   & A &2011-10-08   &  1342230303\\    
     &          &         &     & B &              &  1342230304 \\
     &          &         &     & C &              &  1342230305 \\
\noalign{\smallskip} 

     & o-NH$_2$ & 907.433 & 3b   & A &2012-04-14   & 1342244378 \\   
     &          &         &     & B &            & 1342244379 \\
     &          &         &     & C &            & 1342244380 \\
\noalign{\smallskip} 

& o-NH$_2$ & 952.578 & 2a  & A &2010-04-13       & 1342194706\tablefootmark{b} \\ 
     &          &         &     & B &            & 1342194707\tablefootmark{b} \\
     &          &         &     & C &            & 1342194708\tablefootmark{b} \\
\noalign{\smallskip} 

& p-NH$_2$ & 1\,443.628 & 6a  & A &2012-04-20    & 1342244617 \\   
     &          &         &     & B &            & 1342244618 \\
     &          &         &     & C &            & 1342244619 \\
     &          &         &     & D\tablefootmark{d} &            & 1342244620 \\
     &          &         &     & E &            & 1342244621 \\

\noalign{\smallskip} \noalign{\smallskip} \noalign{\smallskip} 

W31C   & o-NH$_2$ & 648.784 & 2a  & A &2012-04-03  & 1342243675           \\     
            &          &         &     & B &            & 1342243676   \\
            &          &         &     & C &            & 1342243677   \\

\noalign{\smallskip} 
            & o-NH$_2$ & 907.433 & 3b  & A &2011-09-16  & 1342229777\\  
            &          &         &     & B &            & 1342229778  \\
            &          &         &     & C &            & 1342229779  \\
\noalign{\smallskip} 

  & o-NH$_2$ & 952.578 & 3b  & A &2010-03-18  & 1342192319\tablefootmark{b,c} \\ 
            &          &         &     & B &            & 1342192320\tablefootmark{b,c}  \\
            &          &         &     & C &            & 1342192321\tablefootmark{b,c}  \\
\noalign{\smallskip} 
        
& p-NH$_2$ & 1\,443.628 & 6a  & A &2012-04-09    & 1342244072\\   
     &          &         &     & B &            & 1342244073 \\
     &          &         &     & C &            & 1342244074 \\
     &          &         &     & D\tablefootmark{d} &            & 1342244075 \\
     &          &         &     & E\tablefootmark{d} &            & 1342244076 \\

\noalign{\smallskip} \noalign{\smallskip} \noalign{\smallskip} 
W51         & o-NH$_2$ & 648.784 & 2a  & A &2012-04-21  &  1342244806 \\   
            &          &         &     & B &            &  1342244807 \\
            &          &         &     & C &            &  1342244808 \\

\noalign{\smallskip} 
            & o-NH$_2$       & 907.433 & 3b  & A &2012-04-21  & 1342244820 \\ 
            &          &         &     & B &            & 1342244821 \\
            &          &         &     & C &            & 1342244822\\
\noalign{\smallskip} 

  & o-NH$_2$ & 952.578 & 3b  & A &2010-03-18  & 1342192319 \\ 
            &          &         &     & B &            & 1342192320 \\
            &          &         &     & C &            & 1342192321 \\
\noalign{\smallskip} 

& p-NH$_2$ & 1\,443.628 & 6a  & A &2011-11-23    &  1342232678 \\   
     &          &         &     & B &            &  1342232679 \\
     &          &         &     & C &            &  1342232680 \\
     &          &         &     & D &            &  1342232681 \\
     &          &         &     & E &            &  1342232682 \\
 
\noalign{\smallskip} \noalign{\smallskip} \noalign{\smallskip} 

G34.3+0.1   & o-NH$_2$ & 648.384 & 2a  & A &2012-04-21  & 1342244809 \\    
            &          &         &     & B &            & 1342244810  \\
            &          &         &     & C &            & 1342244811   \\
\noalign{\smallskip} 

            & o-NH$_2$ & 907.433 & 3b  & A &2012-03-31  & 1342242871 \\  
            &          &         &     & B &            & 1342242872 \\
            &          &         &     & C &            & 1342242873  \\
\noalign{\smallskip} 

  & o-NH$_2$ & 952.578 & 3b  & A &2010-03-18  & 1342192319 \\ 
            &          &         &     & B &            & 1342192320 \\
            &          &         &     & C &            & 1342192321 \\
\noalign{\smallskip} 

& p-NH$_2$ & 1\,443.628 & 6a  & A &2012-04-20    & 1342244616 \\  
     &          &         &     & B &            & 1342244615 \\
     &          &         &     & C &            & 1342244614 \\
     &          &         &     & D\tablefootmark{d} &            & 1342244613 \\
     &          &         &     & E &            & 1342244612 \\

    \noalign{\smallskip}
\hline 
\label{Table: obsid}
\end{tabular}
\tablefoot{
\tablefoottext{a}{Three different frequency settings of the LO were performed for the three 
ortho transitions and five different LO settings for the para line,  
with approximately 15~km~s$^{-1}$ 
between each setting to determine the  sideband origin of the signals.}  
\tablefoottext{b}{Presented and analysed in paper~II.} 
\tablefoottext{c}{Presented and analysed  in paper~I.} 
\tablefoottext{d}{The H-polarisation was not used in these LO-settings due to strong spikes.} 
}
\end{table*} 
 
 \clearpage

\begin{table}[\!htb] 
\centering
\caption{Hyperfine structure components of $o$-NH$_2$    \mbox{$N_{K_a,K_c}\, J =  2_{1,1} \, 3/2 - 2_{0,2}\, 3/2$}. 
}
\begin{tabular} {cccccc } 
 \hline\hline
     \noalign{\smallskip}

 Frequency\tablefootmark{a}        &A-coeff\tablefootmark{a}& $\Delta v $\tablefootmark{b}      & Rel. Intensity  \\    \noalign{\smallskip}  
(MHz) &(s$^{-1}$)& (km\,s$^{-1}$) & $\frac {A_{ul}\times g_u} {A_{ul}\mathrm{(main)}\times g_u \mathrm{(main)}}$\tablefootmark{c} \\
     \noalign{\smallskip}
     \hline
     \noalign{\smallskip}
   648\,638.83  &  1.52e-4 & 67.19 &  0.003   \\
   648\,647.41 &  1.53e-5  &63.22 &  0.0007  \\
   648\,663.71&  1.07e-3  & 55.69  &  0.023   \\
   648\,667.85 &  2.78e-7 & 53.78  & 0.00001   \\
   648\,673.18&   6.23e-5 & 51.31  &  0.004   \\
   648\,674.50&   3.79e-4 & 50.70 &   0.017  \\
   648\,683.05&   7.19e-4 & 46.75 &   0.016  \\
   648\,691.63 &  2.33e-4 & 42.79 &  0.010   \\
   648\,694.94 &  1.85e-3 & 41.26 &   0.082  \\
   648\,700.27&   3.22e-5 & 38.80 &  0.002   \\
   648\,702.55&   4.60e-3 & 37.74 &  0.202   \\
   648\,704.12 &  1.92e-3 & 37.02 &  0.042   \\
   648\,707.93  & 6.38e-4 & 35.26&   0.014   \\
   648\,708.88  & 1.44e-3 & 34.82 &  0.095   \\
   648\,712.07&  1.52e-3  & 33.34 &   0.067  \\
   648\,712.70 & 2.64e-6  &33.05  &  0.0001   \\
   648\,723.10 & 9.36e-5 & 28.24 &   0.002   \\
   648\,729.00 & 1.50e-4 & 25.52 &   0.003   \\
   648\,729.64&  6.67e-4 & 25.22 &   0.029   \\
   648\,731.69&  2.05e-3 & 24.28 &  0.091  \\
   648\,733.14&  1.22e-3 &  23.61 &  0.054  \\
   648\,735.97&  6.35e-3  & 22.30 &  0.420  \\
   648\,738.47&  2.30e-3 &  21.14 &   0.152  \\
   648\,742.17 & 1.71e-3  & 19.44 &  0.151  \\
   648\,742.33 & 2.52e-3  & 19.36 &   0.166   \\
   648\,746.77 & 5.52e-4   &  17.31 &  0.024   \\
   648\,747.98 & 1.31e-3 &  16.75 &  0.029  \\
   648\,752.12 & 4.19e-5 &  14.83 &   0.002  \\
   648\,754.83&  2.61e-3  & 13.59 &  0.115  \\
   648\,767.85 & 1.10e-3   &  7.57 &  0.049  \\
   648\,771.04&  7.47e-4   &  6.09 &   0.016   \\
   648\,774.17&  6.20e-4  &   4.65 &  0.041  \\
   648\,775.27 & 1.22e-4  &   4.14  &  0.005   \\
   648\,778.03 & 7.67e-4   &  2.86  &  0.051  \\
   648\,779.62 & 1.94e-3   &  2.13  & 0.086  \\
   648\,780.60 & 4.80e-3    & 1.68  & 0.317   \\
   648\,784.23 & 1.14e-2   &   0  &  1.0   \\
   648\,786.83 & 1.55e-4   &  $-$1.20  &  0.007  \\
   648\,795.92 &  4.90e-4  &  $-$5.40  & 0.011  \\
   648\,800.06 &  2.53e-3  &  $-$7.31  &  0.111   \\
   648\,805.39 & 1.22e-3   &  $-$9.78 &  0.080   \\
   648\,809.98 & 1.46e-4   & $-$11.90 &  0.006   \\
   648\,816.30 &  1.64e-3  & $-$14.82 &  0.108   \\
   648\,822.50 & 1.63e-3   & $-$17.68  &  0.144  \\
   648\,834.77 & 1.45e-4   & $-$23.35  &  0.006   \\
   648\,841.09 &  2.17e-4 & $-$26.28 &  0.014  \\

     \noalign{\smallskip}
\hline 
\label{Table: 648 hfs transitions}
\end{tabular}
\tablefoot{
 \tablefoottext{a}{Cologne Database for Molecular Spectroscopy  
\citep[][]{2005JMoSt.742..215M}.} 
\tablefoottext{b}{The velocity offset from the strongest hfs component at  648\,784.228~MHz.} 
\tablefoottext{c}{The sum of the relative intensities  
of the 46 hfs components is 3.87.}   
}
\end{table}

\begin{table}[\!htb] 
\centering
\caption{Hyperfine structure components of $o$-NH$_2$   \mbox{$N_{K_a,K_c}\, J = 2_{0,2}\, 5/2$\,--\,$1_{1,1}\, 3/2$}. 
}
\begin{tabular} {cccccc } 
 \hline\hline
     \noalign{\smallskip}

 Frequency\tablefootmark{a}        &A-coeff\tablefootmark{a}& $\Delta v $\tablefootmark{b}      & Rel. Intensity  \\    \noalign{\smallskip}  
(MHz) &(s$^{-1}$)& (km\,s$^{-1}$) & $\frac {A_{ul}\times g_u} {A_{ul}\mathrm{(main)}\times g_u \mathrm{(main)}}$\tablefootmark{c} \\
     \noalign{\smallskip}
     \hline
     \noalign{\smallskip}
 907\,328.96  &  8.16e-7 & 34.30  & 0.0001   \\
 907\,347.08  &  5.89e-6 & 28.31  & 0.0005   \\
 907\,351.67   & 6.77e-6 & 26.80  & 0.001   \\
 907\,356.75  &  3.11e-6 & 25.12  & 0.0004   \\
 907\,367.88  &  2.62e-5 & 21.44  & 0.004  \\
 907\,378.33  &  4.39e-5 & 17.99  & 0.004  \\
 907\,378.85   & 1.71e-5 & 17.81  & 0.002  \\
 907\,380.30   & 2.30e-5 & 17.34  & 0.002 \\
 907\,380.99   & 1.67e-4 & 17.11  & 0.007   \\
 907\,399.13   & 6.38e-4 & 11.12  & 0.085    \\
 907\,401.56  &  4.80e-4 & 10.31  & 0.085   \\
 907\,401.58  &  9.91e-4 & 10.31  & 0.088   \\
 907\,406.12   & 9.32e-4 &  8.81  & 0.083    \\
 907\,407.48   & 6.05e-4 &  8.36  & 0.081   \\
 907\,414.22  &  2.26e-3 &  6.13  & 0.100    \\
 907\,418.47   & 2.52e-3 &  4.73  & 0.224    \\
 907\,424.18   & 3.88e-3 &  2.84  & 0.517    \\
 907\,426.92   & 3.03e-3 &  1.94  & 0.403    \\
 907\,430.19   & 4.02e-3 &  0.85  & 0.714   \\
 907\,430.27   & 3.89e-3 &  0.83  & 0.691    \\
 907\,432.78   & 4.51e-3 &     0  & 1.0    \\
 907\,432.82  &  2.04e-4 & $-$0.01 &  0.018   \\
 907\,434.80  &  2.20e-3 & $-$0.67 &  0.195    \\
 907\,434.99   & 3.71e-3 & $-$0.73 &  0.494   \\
 907\,440.03  &  6.47e-4 & $-$2.40 &  0.029   \\
 907\,449.02  &  9.17e-5 & $-$5.37 &  0.012    \\
 907\,452.38  &  1.34e-3 & $-$6.48 &  0.059    \\
 907\,456.85  &  1.68e-4 & $-$7.95 &  0.015   \\
 907\,460.62  &  8.71e-4 & $-$9.20 &  0.077    \\
 907\,466.24  &  7.24e-4 &$-$11.05 &  0.096    \\
 907\,472.97   & 1.81e-4 &$-$13.28 &  0.016    \\
 907\,473.55  &  8.11e-4 &$-$13.47 &  0.072   \\
 907\,477.65  &  6.46e-4 &$-$14.82 &  0.086   \\
 907\,480.16   & 5.80e-4 &$-$15.66 &  0.103   \\
 907\,494.03  &  4.36e-5 &$-$20.24 &  0.006   \\
 907\,494.35  &  7.62e-5 &$-$20.34 &  0.010    \\
 907\,507.46  &  9.82e-5 &$-$24.67 &  0.004    \\
 907\,508.79  &  3.31e-5 &$-$25.11 &  0.006   \\
 907\,511.34  &  4.78e-5 &$-$25.96 &  0.004    \\
 907\,516.13  &  2.63e-5 &$-$27.54 &  0.003    \\
 907\,528.04  &  1.42e-5 &$-$31.47 &  0.001    \\
 907\,544.76   & 3.69e-6 &$-$37.00 &  0.0005   \\
 907\,561.46   & 1.98e-7 &$-$42.51 &  0.00003   \\
 
     \noalign{\smallskip}
\hline 
\label{Table: 907 hfs transitions}
\end{tabular}
\tablefoot{
 \tablefoottext{a}{Cologne Database for Molecular Spectroscopy  
\citep[][]{2005JMoSt.742..215M}.} 
\tablefoottext{b}{The velocity offset from the strongest hfs component at  907432.78~MHz.} 
\tablefoottext{c}{The sum of the relative intensities  
of the 43 hfs components is 5.40.}   
}
\end{table} 

\begin{table}[\!htb] 
\centering
\caption{Hyperfine structure components of $o$-NH$_2$   \mbox{$N_{K_a,K_c}\, J = 1_{1,1}\, 3/2$\,--\,$0_{0,0}\, 1/2$}. 
}
\begin{tabular} {cccccc } 
 \hline\hline
     \noalign{\smallskip}

 Frequency\tablefootmark{a}        &A-coef\tablefootmark{a}f& $\Delta v $\tablefootmark{b}      & Rel. Intensity  \\    \noalign{\smallskip}  
(MHz) &(s$^{-1}$)& (km\,s$^{-1}$) & $\frac {A_{ul}\times g_u} {A_{ul}\mathrm{(main)}\times g_u \mathrm{(main)}}$\tablefootmark{c} \\
     \noalign{\smallskip}
     \hline
     \noalign{\smallskip}

   952\,435.66 &   4.86e-6&  44.91  & 0.0002     \\ 
   952\,446.99  & 1.33e-5  &41.34  &   0.0006  \\ 
   952\,463.69  &   3.07e-5 & 36.09  &     0.002   \\ 
   952\,490.73  &   3.5e-3  &  27.58   &    0.079   \\ 
   952\,502.06  &   1.28e-3  & 24.01  &     0.029  \\ 
   952\,503.09  &  1-79e-3  &  23.69   &      0.081  \\ 
   952\,514.42  &  3.65e-3  &  20.12   &     0.164   \\ 
   952\,528.90  &  2.03e-3  & 15.56   &      0.046  \\ 
   952\,533.03  &  2.23e-4  & 14.27   &      0.010   \\ 
   952\,540.23  &  6.53e-3  & 12.00   &     0.147  \\ 
   952\,542.21  &  7.02e-3  & 11.37   &      0.474   \\ 
   952\,549.73  & 2.63e-3   & 9.01    &     0.178  \\ 
   952\,560.41  &   3.32e-3  & 5.65   &     0.149   \\ 
   952\,562.12  &  6.42e-3  & 5.11   &     0.289  \\  
   952\,571.74  &  7.55e-3  & 2.08   &     0.340  \\ 
   952\,573.46  &  3.88e-3  & 1.54   &     0.174\\  
   952\,577.11  &  8.47e-3  & 0.39   &    0.570   \\ 
   952\,578.35  &  1.11e-2     &              0  & 1.0 \\ 
   952\,600.46  & 1.59e-3  &    -6.96   &   0.072\\ 
   952\,615.49  &   3.58e-3  &   -11.69   &     0.081\\ 
   952\,626.82  &  2.73e-3   &   -15.25   &      0.061 \\ 
   952\,627.84  &   2.67e-3   &   -15.57   &     0.120 \\ 
   952\,628.25  &   3.35e-3   &   -15.70   &      0.226 \\ 
   952\,639.17  &  1.40e-3  &   -19.14   &     0.063 \\ 
   952\,653.66  &  9.66e-3   &   -23.70   &     0.022\\ 
   952\,655.64  &  7.43e-4  &    -24.32   &     0.050 \\ 
   952\,659.49  &   4.40e-4  &   -25.54   &     0.020 \\
   952\,664.99  & 1.58e-3   &   -27.27   &   0.036 \\
   952\,686.88  &   2.97e-4  &    -34.16   &     0.013  \\ 
   952\,698.21  &   7.06e-5   &   -37.72  &    0.003 \\

     \noalign{\smallskip}
\hline 
\label{Table: 953 hfs transitions}
\end{tabular}
\tablefoot{ 
 \tablefoottext{a}{Cologne Database for Molecular Spectroscopy  
\citep[][]{2005JMoSt.742..215M}.} 
\tablefoottext{b}The velocity offset from the strongest hfs component at  952578.35\,MHz.
\tablefoottext{c}{The sum of the relative intensities  
of the 30 hfs components is 4.50.}   
}
\end{table} 

\begin{table}[\!htb] 
\centering
\caption{Hyperfine structure components of  $o$-NH$_2$ \mbox{$N_{K_a,K_c}\, J = 4_{2,2}\, 9/2$\,--\,$4_{1,3}\, 9/2$}. 
}
\begin{tabular} {cccccc } 
 \hline\hline
     \noalign{\smallskip}

 Frequency         &A-coeff& $\Delta v $          & Rel. Intensity  \\    \noalign{\smallskip}  
(MHz) &(s$^{-1}$)& (km\,s$^{-1}$) & $\frac {A_{ul}\times g_u} {A_{ul}\mathrm{(main)}\times g_u \mathrm{(main)}}$\tablefootmark{c} \\
     \noalign{\smallskip}
     \hline
     \noalign{\smallskip}
  1012348.20      &      4.86e-7          & 26.04   &0.00003 \\
 1012355.39          &      3.81e-7       &23.91  & 0.00002\\
 1012376.49        &      4.67e-7     &17.66  & 0.00002 \\
 1012383.42        &      1.08e-6      & 15.61  & 0.00004 \\
 1012384.86         &      4.43e-5 &15.18    &0.0025  \\     
 1012391.52     &        6.97e-5 & 13.21  & 0.0033  \\    
 1012391.58      &      4.49e-5  & 13.19   &  0.0021  \\     
 1012392.00        &      4.92e-4 & 13.07 &  0.0329 \\    
 1012398.76       &      5.71e-4 & 11.07  & 0.0327 \\    
 1012398.98       &     7.48e-5 & 11.00  & 0.0029 \\    
 1012399.95      &      4.93e-4 & 10.72 &  0.0282  \\    
 1012405.37       &      5.44e-4& 9.11  & 0.0259  \\    
 1012407.09      &     5.74e-4& 8.60   &  0.0274\\     
 1012412.76       &     6.54e-4 & 6.92   & 0.0250\\    
 1012427.08      &     8.35e-4 & 2.68   & 0.0319 \\   
 1012428.18      &     5.66e-4 & 2.36  & 0.0270 \\     
 1012428.99      &      4.45e-4 & 2.12  & 0.0255 \\
 1012434.50     &      1.11e-3 & 0.49 &  0.0317 \\    
 1012434.59     &       1.44e-2 & 0.46  &  0.6883\\   
 1012435.16    &       7.11e-4 & 0.29 &  0.0271 \\     
 1012435.42     &        1.42e-2 & 0.21  & 0.8112 \\    
 1012435.64      &        5.38e-4 & 0.15  &  0.0257\\     
 1012436.14    &         1.50e-2 & 0 & 1.00 \\ 
 1012442.64      &         1.34e-2 & -1.93 &  0.5099\\      
 1012443.27      &         1.34e-2 & -2.11   & 0.6394\\  
 1012444.02     &          1.44e-2 & -2.33   & 0.8256\\    
 1012448.28      &         1.38e-2 & -3.60 &   0.3962 \\
 1012448.95      &        1.37e-2 & -3.80 &  0.5238\\
 1012449.85     &        1.47e-2 & -4.06 &  0.7034\\  
 1012450.15     &       7.00e-4 & -4.15  & 0.0333 \\  
 1012450.51     &        5.18e-4 & -4.26  & 0.0296  \\      
 1012451.16       &       3.99e-4 & -4.45 &  0.0266 \\  
 1012456.43     &        8.59e-4 & -6.01  & 0.0328 \\    
 1012457.06     &       6.09e-4 & -6.20 &  0.0291 \\
 1012458.22      &       4.63e-9 & -6.54 &  0.0265  \\   
 1012471.24     &        7.14e-9 & -10.40  & 0.0341 \\    
 1012478.82      &        7.83e-4 & -12.64   & 0.0300 \\     
 1012479.55      &        5.92e-4  & -12.86  & 0.0339 \\   
 1012484.47      &        9.07e-4 & -14.31  & 0.0260 \\     
 1012486.33      &        5.82e-6  & -14.86  &   0.0003\\
 1012487.34     &       6.31e-4 & -15.16   & 0.0301  \\    
 1012492.62     &       1.63e-5 & -16.72  &  0.0006 \\ 
 1012493.43      &        6.99e-4 & -16.97 &  0.0267 \\    
 1012494.57      &       3.12e-6 & -17.30  &  0.0002 \\     
 1012501.54      &       8.34e-6 & -19.37 &  0.0004 \\  
     \noalign{\smallskip}
\hline 
\label{Table: 1012 hfs transitions}
\end{tabular}
\tablefoot{ 
 \tablefoottext{a}{Cologne Database for Molecular Spectroscopy  
\citep[][]{2005JMoSt.742..215M}.} 
\tablefoottext{b}The velocity offset from the strongest hfs component at  1012436.1365\,MHz.
\tablefoottext{c}{The sum of the relative intensities  
of the 45 hfs components is 6.81.}   
}
\end{table} 

\begin{table}[\!htb] 
\centering
\caption{Hyperfine structure components of   $p$-NH$_2$ \mbox{$N_{K_a,K_c}\, J = 2_{1,2}\, 5/2$\,--\,$1_{0,1}\, 3/2$}. 
}
\begin{tabular} {cccccc } 
 \hline\hline
     \noalign{\smallskip}

 Frequency         &A-coeff& $\Delta v $          & Rel. Intensity  \\    \noalign{\smallskip}  
(MHz) &(s$^{-1}$)& (km\,s$^{-1}$) & $\frac {A_{ul}\times g_u} {A_{ul}\mathrm{(main)}\times g_u \mathrm{(main)}}$\tablefootmark{c} \\
     \noalign{\smallskip}
     \hline
     \noalign{\smallskip}
1\,442\,539.40   & 1.17e-6  & 226.15   &  1.68e-5      \\
1\,442\,596.40   & 1.15e-6  & 214.31   &  1.65e-5      \\
1\,442\,621.58   & 4.15e-6  & 209.08   &  8.95e-5      \\
1\,443\,564.83   & 3.47e-3  & 13.199   &  4.98e-3      \\
1\,443\,590.01   & 5.56e-3  & 7.9699   &  0.12     \\
1\,443\,595.56   & 8.33e-3  & 6.8177   &  0.12      \\
1\,443\,614.01   & 2.61e-2  & 2.9868   &  0.38      \\
1\,443\,620.75   & 2.92e-2  & 1.5882   &  0.63       \\
1\,443\,628.39   & 3.48e-2  & 0        &  1.0      \\

     \noalign{\smallskip}
\hline 
\label{Table: 1444 hfs transitions}
\end{tabular}
\tablefoot{ 
 \tablefoottext{a}{Cologne Database for Molecular Spectroscopy  
\citep[][]{2005JMoSt.742..215M}.} 
\tablefoottext{b}The velocity offset from the strongest hfs component at   1\,443\,628.39\,MHz.
\tablefoottext{c}{The sum of the relative intensities  
of the 9 hfs components is 2.25.}   
}
\end{table} 

\begin{table}[\!htb] 
\centering
\caption{ALI model properties and resulting excitation temperatures. 
}
\begin{tabular} {lcccc  } 
 \hline\hline
     \noalign{\smallskip}

  & W31C   & W49N & W51 & G34.3 \\
     \noalign{\smallskip}
     \hline
     \noalign{\smallskip}
 \emph{Hot Core} &   \\
 Radius (cm)  & 1.8e18 & 2e18 &   1.45e18 & 9.7e17\\
 $n(\mathrm{H}_2$) (cm$^{-3}$)  & 1.1e5 & 3e5&  3e5 & 1.3e5 \\
 $T_\mathrm{K}$\tablefootmark{a} (K)  & 80 & 76 &  72 & 150 \\
 $\upsilon_\mathrm{turb}$ (km~s$^{-1}$)  & 3.5 & 5.0 & 5 & 3.5  \\
 $X(\mathrm{ortho}$)\tablefootmark{b}   & 5.5e-10 & 8e-11&   1.7e-10 & 2.1e-10\\
 $T_\mathrm{ex}$(ortho-953)  & 18  &26 &  21 & 19\\
 $T_\mathrm{ex}$(para-1444)  & 27  & 39 & 32 & 30 \\
   \noalign{\smallskip}  \noalign{\smallskip} 
 
 \emph{Molecular Envelope} \\
 Radius (cm)  & 1.8e18   & 2e18&  1.45e18   & 1.94e18 \\
 $n(\mathrm{H}_2$) (cm$^{-3}$)  & 2.4e4  & 3e4&  3.5e4   &   2.7e4 \\
 $T_\mathrm{K}$\tablefootmark{a} (K)  & 35 &   40 &  36   &  32\\
 $\upsilon_\mathrm{turb}$ (km~s$^{-1}$)  & 2.6   & 4.0&   5  & 2.7 \\
 $\upsilon_\mathrm{infall}$ (km~s$^{-1}$)  & 3.5 &   2.5&  0  & 2.7 \\
 $X(\mathrm{ortho}$)\tablefootmark{b}   & 1.7e-9 & 1.5e-9 &   1.1e-9&  3.3e-9 \\
 OPR  & 2.6&   5.7\tablefootmark{c} &  3.2\tablefootmark{c} &  4.7\tablefootmark{c}  \\  
 $T_\mathrm{ex}$(ortho-953)  & 12.8  &14 &  13.5 &  10.5 \\
 $T_\mathrm{ex}$(para-1444)  & 17.4  &20.0 &  19.7  &  16.0 \\
     \noalign{\smallskip}
\hline 
\label{table: ali parameters}
\end{tabular}
\tablefoot{ 
 \tablefoottext{a}{$T_\mathrm{dust}  =T_\mathrm{K}$.}   
 \tablefoottext{b}{$X({\rm ortho})$ is the fractional abundance of o-NH$_2$ relative to H$_2$.}
\tablefoottext{c}{The modelling of both the emission and absorption in this source were not successful.}
}
\end{table} 

\end{document}